\documentclass[reqno,12pt,a4paper]{amsart}

\usepackage{mathrsfs}
\usepackage{amssymb}
\usepackage{amsfonts}
\usepackage{latexsym}
\usepackage{amsthm}

\usepackage{graphicx}
\def\lb{\label}

\newcommand{\er}[1]{\textrm{(\ref{#1})}}

\begin{document}


\renewcommand{\theequation}{\arabic{section}.\arabic{equation}}
\theoremstyle{plain}
\newtheorem{theorem}{\bf Theorem}[section]
\newtheorem{lemma}[theorem]{\bf Lemma}
\newtheorem{corollary}[theorem]{\bf Corollary}
\newtheorem{proposition}[theorem]{\bf Proposition}
\newtheorem{definition}[theorem]{\bf Definition}

\def\a{\alpha}  \def\cA{{\mathcal A}}     \def\bA{{\bf A}}  \def\mA{{\mathscr A}}
\def\b{\beta}   \def\cB{{\mathcal B}}     \def\bB{{\bf B}}  \def\mB{{\mathscr B}}
\def\g{\gamma}  \def\cC{{\mathcal C}}     \def\bC{{\bf C}}  \def\mC{{\mathscr C}}
\def\G{\Gamma}  \def\cD{{\mathcal D}}     \def\bD{{\bf D}}  \def\mD{{\mathscr D}}
\def\d{\delta}  \def\cE{{\mathcal E}}     \def\bE{{\bf E}}  \def\mE{{\mathscr E}}
\def\D{\Delta}  \def\cF{{\mathcal F}}     \def\bF{{\bf F}}  \def\mF{{\mathscr F}}
\def\c{\chi}    \def\cG{{\mathcal G}}     \def\bG{{\bf G}}  \def\mG{{\mathscr G}}
\def\z{\zeta}   \def\cH{{\mathcal H}}     \def\bH{{\bf H}}  \def\mH{{\mathscr H}}
\def\e{\eta}    \def\cI{{\mathcal I}}     \def\bI{{\bf I}}  \def\mI{{\mathscr I}}
\def\p{\psi}    \def\cJ{{\mathcal J}}     \def\bJ{{\bf J}}  \def\mJ{{\mathscr J}}
\def\vT{\Theta} \def\cK{{\mathcal K}}     \def\bK{{\bf K}}  \def\mK{{\mathscr K}}
\def\k{\kappa}  \def\cL{{\mathcal L}}     \def\bL{{\bf L}}  \def\mL{{\mathscr L}}
\def\l{\lambda} \def\cM{{\mathcal M}}     \def\bM{{\bf M}}  \def\mM{{\mathscr M}}
\def\L{\Lambda} \def\cN{{\mathcal N}}     \def\bN{{\bf N}}  \def\mN{{\mathscr N}}
\def\m{\mu}     \def\cO{{\mathcal O}}     \def\bO{{\bf O}}  \def\mO{{\mathscr O}}
\def\n{\nu}     \def\cP{{\mathcal P}}     \def\bP{{\bf P}}  \def\mP{{\mathscr P}}
\def\r{\rho}    \def\cQ{{\mathcal Q}}     \def\bQ{{\bf Q}}  \def\mQ{{\mathscr Q}}
\def\s{\sigma}  \def\cR{{\mathcal R}}     \def\bR{{\bf R}}  \def\mR{{\mathscr R}}
\def\vs{\varsigma}  \def\cS{{\mathcal S}}     \def\bS{{\bf S}}  \def\mS{{\mathscr S}}
\def\t{\tau}    \def\cT{{\mathcal T}}     \def\bT{{\bf T}}  \def\mT{{\mathscr T}}
\def\f{\phi}    \def\cU{{\mathcal U}}     \def\bU{{\bf U}}  \def\mU{{\mathscr U}}
\def\F{\Phi}    \def\cV{{\mathcal V}}     \def\bV{{\bf V}}  \def\mV{{\mathscr V}}
\def\P{\Psi}    \def\cW{{\mathcal W}}     \def\bW{{\bf W}}  \def\mW{{\mathscr W}}
\def\o{\omega}  \def\cX{{\mathcal X}}     \def\bX{{\bf X}}  \def\mX{{\mathscr X}}
\def\x{\xi}     \def\cY{{\mathcal Y}}     \def\bY{{\bf Y}}  \def\mY{{\mathscr Y}}
\def\X{\Xi}     \def\cZ{{\mathcal Z}}     \def\bZ{{\bf Z}}  \def\mZ{{\mathscr Z}}
\def\O{\Omega}


\def\mb{{\mathscr b}}
\def\mh{{\mathscr h}}
\def\me{{\mathscr e}}
\def\mk{{\mathscr k}}
\def\mz{{\mathscr z}}
\def\mx{{\mathscr x}}

\def\be{{\bf e}} \def\bc{{\bf c}} \def\bt{{\bf t}}
\def\bx{{\bf x}} \def\by{{\bf y}}
\def\bv{{\bf v}} \def\bu{{\bf u}}
\def\Om{\Omega} \def\bp{{\bf p}}

\newcommand{\gA}{\mathfrak{A}}          \newcommand{\ga}{\mathfrak{a}}
\newcommand{\gB}{\mathfrak{B}}          \newcommand{\gb}{\mathfrak{b}}
\newcommand{\gC}{\mathfrak{C}}          \newcommand{\gc}{\mathfrak{c}}
\newcommand{\gD}{\mathfrak{D}}          \newcommand{\gd}{\mathfrak{d}}
\newcommand{\gE}{\mathfrak{E}}
\newcommand{\gF}{\mathfrak{F}}           \newcommand{\gf}{\mathfrak{f}}
\newcommand{\gG}{\mathfrak{G}}           \newcommand{\gog}{\mathfrak{g}}
\newcommand{\gH}{\mathfrak{H}}           \newcommand{\gh}{\mathfrak{h}}
\newcommand{\gI}{\mathfrak{I}}           \newcommand{\gi}{\mathfrak{i}}
\newcommand{\gJ}{\mathfrak{J}}           \newcommand{\gj}{\mathfrak{j}}
\newcommand{\gK}{\mathfrak{K}}            \newcommand{\gk}{\mathfrak{k}}
\newcommand{\gL}{\mathfrak{L}}            \newcommand{\gl}{\mathfrak{l}}
\newcommand{\gM}{\mathfrak{M}}            \newcommand{\gm}{\mathfrak{m}}
\newcommand{\gN}{\mathfrak{N}}            \newcommand{\gn}{\mathfrak{n}}
\newcommand{\gO}{\mathfrak{O}}
\newcommand{\gP}{\mathfrak{P}}             \newcommand{\gp}{\mathfrak{p}}
\newcommand{\gQ}{\mathfrak{Q}}             \newcommand{\gq}{\mathfrak{q}}
\newcommand{\gR}{\mathfrak{R}}             \newcommand{\gr}{\mathfrak{r}}
\newcommand{\gS}{\mathfrak{S}}              \newcommand{\gs}{\mathfrak{s}}
\newcommand{\gT}{\mathfrak{T}}             \newcommand{\gt}{\mathfrak{t}}
\newcommand{\gU}{\mathfrak{U}}             \newcommand{\gu}{\mathfrak{u}}
\newcommand{\gV}{\mathfrak{V}}             \newcommand{\gv}{\mathfrak{v}}
\newcommand{\gW}{\mathfrak{W}}             \newcommand{\gw}{\mathfrak{w}}
\newcommand{\gX}{\mathfrak{X}}               \newcommand{\gx}{\mathfrak{x}}
\newcommand{\gY}{\mathfrak{Y}}              \newcommand{\gy}{\mathfrak{y}}
\newcommand{\gZ}{\mathfrak{Z}}             \newcommand{\gz}{\mathfrak{z}}

\def\ve{\varepsilon} \def\vt{\vartheta} \def\vp{\varphi}  \def\vk{\varkappa}
\def\vr{\varrho}

\def\Z{{\mathbb Z}} \def\R{{\mathbb R}} \def\C{{\mathbb C}}  \def\K{{\mathbb K}}
\def\T{{\mathbb T}} \def\N{{\mathbb N}} \def\dD{{\mathbb D}} \def\S{{\mathbb S}}
\def\B{{\mathbb B}}


\def\la{\leftarrow}              \def\ra{\rightarrow}     \def\Ra{\Rightarrow}
\def\ua{\uparrow}                \def\da{\downarrow}
\def\lra{\leftrightarrow}        \def\Lra{\Leftrightarrow}
\newcommand{\abs}[1]{\lvert#1\rvert}
\newcommand{\br}[1]{\left(#1\right)}

\def\lan{\langle} \def\ran{\rangle}


\def\lt{\biggl}                  \def\rt{\biggr}
\def\ol{\overline}               \def\wt{\widetilde}
\def\no{\noindent}


\let\ge\geqslant                 \let\le\leqslant
\def\lan{\langle}                \def\ran{\rangle}
\def\/{\over}                    \def\iy{\infty}
\def\sm{\setminus}               \def\es{\emptyset}
\def\ss{\subset}                 \def\ts{\times}
\def\pa{\partial}                \def\os{\oplus}
\def\om{\ominus}                 \def\ev{\equiv}
\def\iint{\int\!\!\!\int}        \def\iintt{\mathop{\int\!\!\int\!\!\dots\!\!\int}\limits}
\def\el2{\ell^{\,2}}             \def\1{1\!\!1}
\def\sh{\sharp}
\def\wh{\widehat}
\def\bs{\backslash}
\def\na{\nabla}
\def\ti{\tilde}
\def\hb{\hbar}
\def\cd{\centerdot}

\def\sh{\mathop{\mathrm{sh}}\nolimits}
\def\all{\mathop{\mathrm{all}}\nolimits}
\def\Area{\mathop{\mathrm{Area}}\nolimits}
\def\arg{\mathop{\mathrm{arg}}\nolimits}
\def\const{\mathop{\mathrm{const}}\nolimits}
\def\det{\mathop{\mathrm{det}}\nolimits}
\def\diag{\mathop{\mathrm{diag}}\nolimits}
\def\diam{\mathop{\mathrm{diam}}\nolimits}
\def\dim{\mathop{\mathrm{dim}}\nolimits}
\def\dist{\mathop{\mathrm{dist}}\nolimits}
\def\Im{\mathop{\mathrm{Im}}\nolimits}
\def\Iso{\mathop{\mathrm{Iso}}\nolimits}
\def\Ker{\mathop{\mathrm{Ker}}\nolimits}
\def\Lip{\mathop{\mathrm{Lip}}\nolimits}
\def\rank{\mathop{\mathrm{rank}}\limits}
\def\Ran{\mathop{\mathrm{Ran}}\nolimits}
\def\Re{\mathop{\mathrm{Re}}\nolimits}
\def\Res{\mathop{\mathrm{Res}}\nolimits}
\def\res{\mathop{\mathrm{res}}\limits}
\def\sign{\mathop{\mathrm{sign}}\nolimits}
\def\span{\mathop{\mathrm{span}}\nolimits}
\def\supp{\mathop{\mathrm{supp}}\nolimits}
\def\Tr{\mathop{\mathrm{Tr}}\nolimits}
\def\BBox{\hspace{1mm}\vrule height6pt width5.5pt depth0pt \hspace{6pt}}
\def\where{\mathop{\mathrm{where}}\nolimits}
\def\as{\mathop{\mathrm{as}}\nolimits}



\newcommand\nh[2]{\widehat{#1}\vphantom{#1}^{(#2)}}
\def\dia{\diamond}

\def\Oplus{\bigoplus\nolimits}



\def\qqq{\qquad}
\def\qq{\quad}
\let\ge\geqslant
\let\le\leqslant
\let\geq\geqslant
\let\leq\leqslant
\newcommand{\ca}{\begin{cases}}
\newcommand{\ac}{\end{cases}}
\newcommand{\ma}{\begin{pmatrix}}
\newcommand{\am}{\end{pmatrix}}
\renewcommand{\[}{\begin{equation}}
\renewcommand{\]}{\end{equation}}
\def\eq{\begin{equation}}
\def\qe{\end{equation}}
\def\[{\begin{equation}}
\def\bu{\bullet}
\def\ced{\centerdot}
\def\tes{\textstyle}


\title[{Isomorphic inverse problems }]
 {Isomorphic inverse problems}


\date{\today}
\author[Evgeny Korotyaev]{Evgeny L. Korotyaev}
\address{Department of Math. Analysis, Saint-Petersburg State University,
Universitetskaya nab. 7/9, St. Petersburg, 199034, Russia, and HSE
University, 3A Kantemirovskaya ulitsa, St. Petersburg, 194100,
Russia
 \ korotyaev@gmail.com, \
e.korotyaev@spbu.ru}


\begin{abstract}
 Consider two inverse problems for Sturm-Liouville  problems on the
 unit interval. It means that there
 are two corresponding mappings
$F, f$ from a Hilbert space of potentials   $H$ into their spectral
data. They are called isomorphic if $F$ is a composition of $f$ and
some isomorphism  $U$ of $H$ onto itself. A isomorphic class is a
collection of inverse problems isomorphic  to each other. We
consider basic Sturm-Liouville problems on the unit interval and on
the circle and describe their isomorphic classes of inverse
problems. For example, we prove that the inverse problems  for the
case of Dirichlet and Neumann boundary conditions are isomorphic.
The proof is based on the non-linear analysis.

\end{abstract}

\maketitle



\section {Introduction and main results}
\setcounter{equation}{0}

\subsection{Introduction }

We consider the Sturm-Liouville  problems $ -y''+qy=\l y$ on the
unit interval $ [0,1]$ under basic boundary conditions  or on the
unit circle $\S^1$. Here the  potential $q$ is  real and belongs to
the space $L^2(0,1)$. There are a lot of results about the inverse
problems for the Sturm-Liouville operators on $ [0,1]$ and on
$\S^1$.  These inverse problems   were investigated by many authors
(G.~Borg, I.~M.~Gel'fand, B.~M.~Levitan, V.~A.~Marchenko,
E.~Trubowitz, ..), see  the monographs \cite{L87}, \cite{M86},
\cite{PT87} and references therein. In general, the study of an
inverse spectral problem consists of the following parts:

\noindent (i) Uniqueness: prove that the spectral data
(eigenvalues plus some additional parameters) determine the potential uniquely);\\
(ii) Reconstruction: reconstruct the potential from spectral data;\\
(iii) Characterization: describe all spectral data corresponding to
fixed classes of potentials. \\
(iv) Stability estimates: obtain a priori two sided estimates of the
potential and spectral data.


We will discuss their additional {\it isomorphic} properties.

\no {\bf Definition.}{\it Let $f$ and $f_o$ be mappings from a
Hilbert space $\cK$ to a set $X$. They are called isomorphic if
$f_o=f\circ U$ for some isomorphism (in general, non-linear) $U$ of
$\cK$ onto itself.}

Note that if some of them is a bijection, then $U$ is a unique
canonical automorphism of $\cK$.

 Consider two inverse problems for
Sturm-Liouville  problems, when potentials $q$ belong to a
corresponding Hilbert space $\cK$. Thus there are two mappings
$f:\cK\to X$ and $f_o:\cK\to X$, where $X$ is their set of spectral
data. They are called isomorphic if $f_o=f\circ U$ for some
isomorphism (in general, it is non-linear) $U$ of $\cK$ onto itself.
If  $U$ is an unitary linear operator, then these two inverse
problems are called unitarily equivalent.

Note that if two Sturm-Liouville inverse problems are isomorphic,
then we have

1) If the first has some property from (i)-(iv), then the second
also has it. For example, the first has uniqueness iff the second
has uniqueness.

2) Eigenvalues of the first problem  have some asymptotics for each
potential iff
 eigenvalues of the second  problem have similar asymptotics.

 3) The first problem  has some trace formula iff
 the second  problem  has a similar trace formula.

We write our main results:

A) We describe all isomorphic Sturm-Liouville inverse problems on
the unit interval under the basic boundary conditions and on the
circle. The corresponding automorphism $U$ is obtained in explicit
form.

B)  The same is made for potentials from Sobolev spaces. The needed
new sharp asymptotics of norming constants are determined.

To the best of our knowledge the obtained results have no analogies
in existing literature. Our proof uses observations 1)-3) and also
the following results and methods:

$\cd$ two spectra mapping (Marchenko-Ostrovski \cite{MO75}),

$\cd$  the Dirichlet eigenvalues and norming constants mapping
(P\"oschel-Trubowitz \cite{PT87}),

$\cd$ the four  spectra mapping (Korotyaev \cite{K19}),

$\cd$ results  of Marchenko--Ostrovski \cite{MO75} and Korotyaev
\cite{K99} about inverse periodic problems.

We consider four Sturm-Liouville problems on the interval $[0,1]$
with the Dirichlet and Neumann boundary conditions:
\[
\lb{DN} -f''+qf=\l f,\qqq
\begin{aligned}
&     \qqq f(0)= \textstyle f(1)=0, \qq
\\
&   \qqq  f'(0)=\textstyle f'(1)=0,
\end{aligned}
\]
and with the so-called mixed boundary
conditions:
\[
\lb{mbc} -f''+qf=\l f,\qqq
\begin{aligned}
&\qqq f(0)=\textstyle f'(1)=0, \qq
\\
& \qqq f'(0)=\textstyle f(1)=0,
\end{aligned}
\]
where $\l\in \C$.
 Here the potential $q$ belongs to the real Hilbert space $\cL$ defined by
$$
  \cL=\rt\{q\in L^2([0,1],\R): \ \int_0^1
qdx=0\rt\}
$$
equipped with the norm $\|q\|^2=\int_0^1 q^2(x)dx$. There are a lot
of results about these problems, see, e.g., the books \cite{L87},
\cite{M86}, \cite{PT87}.  Let $\m_n$ and $\n_0, \n_n, n\ge 1$ be
eigenvalues of the Dirichlet and Neumann problems respectively. Let
$\t_n,$ and $\vr_n, n\ge 1$ be eigenvalues of the first and the
second problem  respectively with mixed boundary conditions
\er{mbc}, and we say shortly mixed eigenvalues.  All these
eigenvalues are simple and satisfy
\[
\lb{bx}
\begin{aligned}
& \n_0<\ol {\t_1, \vr_1}<\ol {\m_1,\n_1}<\ol {\t_2, \vr_2}< \ol
{\m_2,\n_2}<...,
\\
& \n_n,\m_n=\m_n^o+o(1),\qq \t_n, \vr_n=\t_n^o+o(1)\qq \as \ n\to
\iy,
\end{aligned}
\]
where $\ol {u,v}$ denotes $\min \{u,v\}\le \max \{u,v\}$ for
shortness, and $\n_0^o=0, \n_n^o=\m_n^o=(\pi n)^2$ and $
\t_n^o=\vr_n^o=\pi^2 (n-{1\/2})^2, n\ge 1$ are the corresponding
unperturbed eigenvalues.

 We introduce  the fundamental solutions
$\vp(x,\l), \vt (x,\l)$ of the equation
$$
  -f''+q(x)f=\l f,\ \ \  \ \ \ \l\in \C,
$$
under conditions: $\vp'(0,\l)=\vt (0,\l)=1$ and
$\vp(0,\l)=\vt'(0,\l)=0$. Here and below $(\,')={\pa\/\pa x}$ and
$(\dot{{\,}})={\pa\/\pa \l}$. Note that $\{\m_n\}, \{\n_n\},
\{\t_n\}$ and $\{\vr_n\}$ are zeros of $\vp (1,\l), \vt'(1,\l),
\vp'(1,\l)$ and $\vt(1,\l)$ respectively. Introduce the real Hilbert
spaces $\ell_k^2=\ell_k^2(\N), k\in\R$ of real sequences
$v=(v_n)_1^\iy$ equipped with the norm
$$
\|v\|_{k}^2=\sum _{n\ge 1 }(2\pi n)^{2k}v_n^2,\qq {\rm  and\ let}
\qq \ell^2=\ell_0^2, \qq \|\cdot \|=\|\cdot \|_0.
$$
Following the book of P\"oschel and Trubowitz \cite{PT87} we define
a set $\gJ^o$  of all real, strictly increasing sequences by
$$
\gJ^o=\Big\{ s\!=\!(s_n)_1^\iy: s_1<s_2<....., \qq
s_n\!=\!\m_n^o\!+\wt s_{n\,},\qqq \wt s=(\wt
s_{n\,})_1^\iy\!\in\!\ell^2     \Big\}.
$$
 Note that the mapping
$s\lra \wt s $ is a natural coordinate map between $\gJ^o$ and some
open convex subset $\wt \gJ^o=\Big\{ \wt s\!=\!(\wt
s_n)_1^\iy\in\!\ell^2: \m_1^o+\wt s_{1\,}<\m_2^o+\wt s_{2\,}<.....
\Big\} $ of $\ell^2$\,. Following  \cite{PT87} we identify $\gJ^o$
and $\wt \gJ^o$ using this mapping. As in \cite{PT87} this
identification allows to do analysis on $\gJ^o$ as if it was an open
convex subset of $\ell^2$. We also define similar sets $\gJ^1$ and
$\gJ$ of all real, strictly increasing sequences by
$$
\begin{aligned}
\gJ^1=\Big\{ t\!=\!(t_n)_1^\iy: t_1<t_2<....., \qq
t_n\!=\!\t_n^o\!+\wt t_{n\,},\qqq \wt t=(\wt
t_{n\,})_1^\iy\!\in\!\ell^2     \Big\},
\\
\textstyle \gJ=\Big\{ t\!=\!(t_n)_1^\iy: t_1<t_2<....., \qq
t_n\!=({\pi n\/2})^2+\wt t_{n\,},\qqq \wt t=(\wt
t_{n\,})_1^\iy\!\in\!\ell^2 \Big\}.
\end{aligned}
$$
Introduce  1-spectra mappings $\m$ and $\n$ from $\cL$ into $\gJ^o$
and $\t$ and $\vr$ from $\cL$ into $\gJ^1$ by
\[
\lb{mntz1}
\begin{aligned}
q\to \m=(\m_n)_1^\iy,\ \qq  q\to \n=(\n_n)_1^\iy,
\\
 q\to \t=(\t_n)_1^\iy,\ \qq q\to \vr=(\vr_n)_1^\iy.
 \end{aligned}
\]
For two 1-spectra mappings (only for strongly increasing and
alternate) we construct standard 2-spectra mappings of strongly
increasing sequences. For example, for $\t=(\t_n)_1^\iy\in \gJ^1$
and $\m=(\m_n)_1^\iy\in \gJ^o$ such that $\t_1<\m_1<\t_2<\m_2<...$
we define a 2-spectra mapping $\t \star \m$ from $\cL$ into $\gJ$ as
\[
\lb{st} q\to \t \star \m=(\t_1,\m_1,\t_2,\m_,....).
\]
The mapping $\t\star \m$ is a bijection between $\cL$ and $\gJ$. It
is a classical result of Marchenko and Ostrovski \cite{MO75}.

Following Trubowitz and co-authors \cite{IT83}, \cite{DT84} we
introduce norming constants  $h_{s,n}, \gh_{s,n}$ (associated with
the Dirichlet and Neumann eigenvalues) and the corresponding
mappings by
\[
\lb{mntz2}
\begin{aligned}
h_{s,n}=\ln |\vp'(1,\m_n)|, \qqq &  \gh_{s,n}=\ln |\vt(1,\n_n)|,\qqq
n\ge 1,\qq \gh_{s,0}=\ln |\vt(1,\n_0)|,
\\
   q\to h_{(s)}=(h_{s,n})_{n=1}^\iy,\qqq & q\to
\gh_{(s)}=(\gh_{s,n})_{n=1}^\iy.
 \end{aligned}
\]
It is known that the  mappings $\m\ts h_{(s)}$ is a bijections
between $\cL$ and $\gJ^o\ts \ell_1^2$. This is a classical result of
P\"oschel and Trubowitz \cite{PT87}.

 We sometimes write $\m_n(q), \n_n(q),...$
instead of $\m_n, \n_n,...$,  when several potentials  are being
dealt with. We discuss a mapping $q\to \m\ts (D_n)_1^\iy$ introduce
by Marchenko \cite{M50}, where $D_n$ is the normalizing constant
associated with Dirichlet eigenvalue $\m_n$ and defined by
\[
\lb{a1} \tes D_n(q)=\int_0^1\vp^2(x,\m_n(q),q)dx,\qq n\in \N, \qqq
\where \ D_n(0)={1\/2\m_n^o}.
\]
Marchenko \cite{M50} proved that the spectral  data  $\m_n, D_n,
n\in \N$ determines the potential uniquely. Gel'fand and Levitan
\cite{GL51} created a basic method to reconstruct the potential $q$
from $\m_n, D_n, n\in \N$: they determined an integral equation and
expressed $q$ in terms of its solution.
 It is convenient to modify constants $D_n$ and
define another mapping $q\to \a=(\a_n)_1^\iy$, where the components
$\a_n$ are given by
\[
\lb{a2} \a_n=\log {D_n(q)\/D_n(0)}=\log \big[2\m_n^o D_n(q)\big],
\qqq \qq n\in \N.
\]

We discuss similar mapping $q\to \n\ts (N_n)_1^\iy$, where $N_n$ is
a normalizing constant, associated with Neumann eigenvalue $\n_n$
and defined by
\[
\lb{n1} N_n(q)=\int_0^1\vt^2(x,\n_n,q)dx,\qq n\in \N, \qqq \where \
\tes N_n(0)={1\/2}.
\]
It is more convenient to modify constants $N_n$ and to define
another mapping $q\to \b=(\b_n)_1^\iy$, where the components $\b_n$
are given by
\[
\lb{n2} \b_n=\ln [2N_n(q)].
\]

 Recall some
definitions. We write $\cK_\C$ for the complexification of the real
Hilbert space $\cK$. Suppose that $ \cK, \cS$ are real separable
Hilbert spaces. The mapping $f:\cK\to \cS$ is a local real analytic
isomorphism iff for any $y\!\in\!\cK$ it has an analytic
continuation $\wt{f}$ into some complex neighborhood
$\!V\!\ss\!{\cK}_\C$ of $ y$, which is a bijection between $V$ and
some open set $\wt{f}(V)\!\ss\!{\cS}_\C$ and if $\wt{f}$,
$\wt{f}^{-1}$ are analytic mappings on $V$\,, $\wt{f}(V)$
respectively. The mapping $f$ is a real-analytic bijection (shortly
a RAB) between $\cK$ and $\cS$ if it is both a bijection and a local
real analytic isomorphism.


\subsection{Short review}

 We shortly discuss  well-known results about
inverse  problems  for  Sturm-Liouville operators on the unit
interval under different boundary conditions, which are used in our
paper. We recall only some important steps mainly  on the
characterization problem. Marchenko and Ostrovski in \cite{MO75}
solved the inverse problem for the 2-spectra mapping $\t\star \m$.
Their proof is based on the inverse scattering on the half line
(with decreasing potentials) and sharp asymptotics for eigenvalues
$\m_n, \t_n$.

Trubowitz and co-authors (\cite{DT84}, \cite{IT83}, \cite{ IMT84},
\cite{PT87}) suggested an analytic approach. It is based on
analyticity of the mapping
$\{\mathrm{potentials}\}\mapsto\{\mathrm{spectral\ data}\}$ and an
explicit reconstruction procedure for the special case when only one
spectral parameter has been changed.  The excellent  book
\cite{PT87} is devoted to the mapping $q \to \m\star h_{(s)}$, where
the inverse problem is solved, including the characterization of the
spectral data.
 Also, this approach was applied to
other inverse problems with purely discrete spectrum: for impedance
\cite{CM93}, \cite{CM93a}, \cite{K00};  singular Sturm-Liouville
operators on a finite interval \cite{GR88}; periodic potentials
\cite{GT87}, \cite{KK97}, \cite{K99}, perturbed harmonic oscillators
\cite{CKK04}, vector-valued operators \cite{CK09}, and Birkhoff
coordinates for the KdV equation on the circle \cite{KP03}, and see
references therein.

Now we discuss the periodic case. Note that only the periodic
eigenvalues do not determine a potential uniquely and we need to add
auxiliary spectral  data: Dirichlet (or Neumann) eigenvalues plus
sequence of signs $\pm$. There are only two mappings related to the
characterization for inverse periodic problems:

1) in terms of local maxima and minima of the Lyapunov functions on
the real line,

2) in terms of gaps.

\no Marchenko and Ostrovski \cite{MO75} solved the inverse problems
(including characterization and stability estimates) in terms of the
local maxima and minima of Lyapunov functions on the real line. The
proof is based on the inverse scattering on the half line (with
decreasing potentials), sharp asymptotics of periodic eigenvalues
and new results about conformal mappings for the  quasimomentum. A
shorter proof was given in \cite{K97}. Korotyaev \cite{K99},
\cite{K98} solved the inverse problems  in terms of gap lengths. The
proof (including characterization) was based on analyticity of the
mapping and a priori estimate of potentials in terms of gap lengths
from \cite{K98}.

\subsection{Main results on the unit interval}
In order  to discuss main results we introduce our basic
transformations. We define the 4-spectra mapping $\gf: \cL\to
\ell^2$ from a recent paper \cite{K19} by
\[
\lb{dgf} q\to \gf(q)=(\gf_n(q))_1^{\iy },\qq \gf_{2n-1}=\vr_n
-\t_n,\qqq  \gf_{2n}=\n_n -\m_n, \ \ \ \ \ n\ge 1,
\]
which will be basic for us. Recall a result from \cite{K19}:
 {\it the 4-spectra mapping $\gf: \cL\to \ell^2$    defined by
\er{dgf} is a RAB between $\cL$ and $\ell^2$}, see more in Theorem
\ref{TMO}. Let $\gS$ be a set of all diagonal operators $\s= \diag
(\s_1,\s_2,...)$ on $\ell^2$, or shortly $\s=(\s_j)_1^\iy$, where
$\s_j\in \{\pm 1\}, j\in \N$. This set $\gS$ defines so-called the
lamplighter group, see \cite{CM17}. For each $\s\in \gS$ we define a
mapping $ \cU_\s$ by
\[
\lb{dj}
 \cU_\s=\gf^{-1} \s\gf: \cL\to
\cL,
\]
where $\gf$ is given by \er{dgf}.  These mappings have the following
properties:

\begin{proposition}
\lb{TPr} Let $\cU_\s, \s\in \gS$ be defined by \er{dj}. Then each
$\cU_\s$  is a RAB of $\cL$ onto itself and  satisfies
\[
\lb{dUx}
\begin{aligned}
 \cU_\s=\cU_\s^{-1}, \qqq \cU_\s\ \cU_{\s'}=\cU_{\s\s'}=\cU_{\s'}\
\cU_{\s},\qqq \forall \qq \s, \s'\in \gS,
\end{aligned}
\]
\[
\lb{dUq} \|\cU_\s(q)\|=\|q\|\qqq \forall \ q\in \cL.
\]
\end{proposition}

\no {\bf Remark.} 1) The mapping $\gf$ is non-linear, but $\cU_\s$
keeps the norm on $\cL$, see \er{dUq}.

\no 2) To prove Proposition \ref{TPr} we use the bijection of the
mapping  $\gf$ from \cite{K19}. Its proof is based on the bijections
of the mapping $\t\star \m$ and on inverse problems  for periodic
potentials \cite{K99}.

  Define two specific  mappings, often used in our paper:
\[
\lb{dU}
\begin{aligned}
& U_1=\cU_\s,\qq \where \qq \s= (\s_{j})_1^\iy, \ \s_{j}=-1 \qq \ \
\forall \ j\in \N,
\\
& U_o=\cU_\s,\qq \where \qq \s=(\s_{j})_1^\iy, \ \s_{j}= (-1)^j\ \
\forall \ j\in \N.
\end{aligned}
\]
Below we show that $U_o=\cR$, where $\cR$ is  a reflection (unitary)
operator on $\cL$  given  by $(\cR y)(x)=y(1-x), x\in (0,1)$. For
each $\gn=(\gn_n)_1^\iy\in \gJ$ we define a potential $\gq$ by
\[
\lb{rq1} \tes  \gq(x)=-2{d^2\/dx^2} \log
\Big(\G_\gn\det\O(x,\gn)\Big),\ \ \ x\in (0,1),
\]
where $\O(x,\gn), x\in (0,1)$ is the infinite matrix whose elements
$\O_{n,j}(x,\gn)$ are given by
\[
\lb{rq2}
\begin{aligned}
 \O_{n,j}(x,\gn)={\gn_n-\gn_n^o \/ \gn_n-\gn_j^o} \lt\{\cos
\sqrt{\gn_n}x+{(-1)^n-\cos 2\sqrt{\gn_n}\/\sin 2\sqrt{\gn_n}}\sin
\sqrt{\gn_n} x, {\sin {\pi j\/2} x\/{\pi j\/2}}\rt\}_{w},
\end{aligned}
\]
where   $\{u,v\}_{w}=uv'-u'v$  and
\[
\lb{rq3} \G_\gn=\prod_{j>n\ge 1}\lt({\gn_n-\gn_n^o
\/\gn_n-\gn_j}\cdot {\gn_n^o-\gn_j \/  \gn_n^o-\gn_j^o}\rt),\qqq
\gn_n^o=\Big({\pi n\/2}\Big)^2.
\]
 Here $\O-I$ is a trace class operator, so
$\det\O(x,p)$ is well defined. A recovering of a potential $q$ by
\er{rq1} via the 2-spectra mapping
 $\gn=\t\star \m,\qq \gn_{2j-1}=\t_j,\qq \gn_{2j}=\m_j, j\in \N$,
was performed  in \cite{K19} (see also \cite{PT87} for even
potentials). The first main result describes equivalent inverse
problems on the unit interval.

 \begin{theorem}
\lb{T1}

 Let mappings $U_0, U_1$ be defined  by \er{dU}. Then 1-spectra
mappings $\m=(\m_n)_1^\iy,  \n=(\n_n)_1^\iy,$ and $\t=(\t_n)_1^\iy,
\vr=(\vr_n)_1^\iy$  defined by \er{mntz1} satisfies

\no i) All 2-spectra mappings $\t\star \m, \vr\star \n$,  $\vr\star
\m$ and $\t\star \n$ acting from $\cL$ into $\gJ$  are isomorphic.
Moreover, each of them  is a RAB between $\cL$ and $\gJ$ and
satisfies
\[
\lb{enm}
\begin{aligned}
\t\star \m=(\vr\star \n)\circ U_1=(\vr\star \m)\circ U_o=(\t\star
\n)\circ U_o\circ U_1.
\end{aligned}
\]
\no ii)   Let $q\in \cL$. Then a potential $U_1(q)$ is given by
\er{rq1}, where $\gn=\vr\star \n$.

\no iii) Let $q\in \cL$.  Then a potential $(U_0U_1)(q)$ is given by
\er{rq1}, where $\gn=\t\star \n$.

\end{theorem}

\no {\bf Remark.} 1) In the proof we show \er{enm}, then we obtain
i), since the mapping $\t\star \m$ is a bijection between $\cL$ and
$\gJ$ \cite{MO75} and $U_0, U_1$ are bijections of $\cL$ onto
itself.

\no 2) We connect all 2-spectra mappings by the bijections  $U_o,
U_1$ and $U_oU_1$.

\medskip

We discuss the mapping $q\to $ eigenvalues plus norming (or
normalizing) constants.

 \begin{theorem}
\lb{T1x}
i)  Mappings $\m\ts h_{(s)}, \n \ts \gh_{(s)}$,  $\m \ts \gh_{(s)}$
and $\n \ts h_{(s)}$ (defined by  \er{mntz1}, \er{mntz2}) are
isomorphic as mappings from $\cL$ into $\gJ^o\ts \ell_1^2$ and
satisfy
\[
\lb{enm1} \begin{aligned} \m\ts h_{(s)}=(\n \ts \gh_{(s)})\circ
U_1=(\m \ts \gh_{(s)})\circ U_o=(\n \ts h_{(s)})\circ U_o\circ U_1.
\end{aligned}
\]
 Moreover, each of them is a RAB between $\cL$ and $\gJ^o\ts
\ell_1^2$.

\no ii) Each of the mappings $\m\ts \a$ and $\n \ts \b$  (defined by
\er{a2}, \er{n2}) acting from $\cL$ into $\gJ^o\ts \ell_1^2$ is a
bijection between $\cL$ and $\gJ^o\ts \ell_1^2$  and the following
trace formula holds true
\[
\lb{trN1}  {1\/N_0}-1=\sum_{n\ge 1} \Big(2-{1\/N_n}\Big),
\]
where the series converges absolutely and uniformly on bounded
subsets of $\cL$.

\no iii) Define another normalizing mapping $q\to \hat \b=
(\hat\b_n)_1^\iy$, where  $\hat \b_n=\b_n-\ln {\n_n-\n_0\/\n_n^o}$.
Then the mapping $\n \ts \hat \b$ is a bijection between $\cL$ and
$\gJ^o\ts \ell_1^2$ and satisfies
\[
\lb{nc1q} \begin{aligned} \m\ts \a=(\n \ts \hat \b)\circ U_1.
\end{aligned}
\]

 \end{theorem}

\no {\bf Remark.} 1) The  mapping $\n \ts \gh_{(s)}$ is a bijection
between $\cL$ and $\gJ$, see  \cite{KC09}. The mapping $\n \ts
\gh_{(s)}$ does not use  the eigenvalue $\n_0$ and the norming
constant $\gh_{s,0}$ since $\n_n, \gh_{s,n}, n\ge 0$ are dependent,
see  trace formulas \er{i1t}, \er{trN1}.

\no 2)  In order to prove that the mappings in ii)  are bijections
we show, that they are isomorphic to the mapping $\m\ts h_{(s)}$,
which is well studied in \cite{PT87}. Thus the proof is short.

\medskip

\no {\bf Final Remarks.}  Below we discuss also following isomorphic
inverse problems:

\no $\bu$ The case of mixed eigenvalues is considered in Sect. 4.

\no $\bu$ The case of smooth potentials is considered in Sect. 5.
The needed new sharp asymptotics for the norming constants for
potentials from Sobolev space are obtained in Sect. 7.

\no $\bu$  The case of periodic potentials is discussed in Sect. 6
(including stability eliminates).

In Sect. 2  we prove preliminary results about the mappings $\cU_\s,
U_1, U_o$.
 In Sect. 3  the main results for Dirichlet and Neumann b.c.
are proved.  In Sect. 7 we determined some specific  asymptotics for
the fundamental solutions, and trace formulas.

\section {Properties of $\cU_\s, U_1, U_o$}
\setcounter{equation}{0}

\subsection{Preliminaries }
We introduce mappings $q\to \gt=(\gt_n)_1^\iy$ and $ q\to
\gr=(\gr_n)_1^\iy$, where the components are norming constants
$\gt_n, \gr_n$ (associated with the mixed eigenvalues $\t_n, \vr_n$,
respectively) and are given by
\[
\label{nce4}
\begin{aligned}
\gt_n=\ln |\vp(1,\t_n)\sqrt{\t_n^o}|,& \qqq \gr_n(q)=-\ln
|\vt'(1,\vr_n)/\sqrt{\t_n^o}|=\ln |\vp(1,\vr_n)\sqrt{\vr_n^o}|,
\end{aligned}
\]
where $\vt'(1,\vr_n)\vp(1,\vr_n)=-1$ has been used. Note that
$\gt(q)\in \ell_1^2$, see \cite{KC09}. In the case of   mix boundary
conditions $y(0)\!=\!0$, $y'(1)\!=\!0$ the spectral data
$(\t_n)_{1}^{\iy}$, $(\gt_n)_{1}^{\iy}$ are not independent since
they satisfy the trace formula \er{b1x}.  It turns out that the
first eigenvalue $\t_1$ can be uniquely reconstructed from the other
spectral data $(\t_n)_{2}^{\iy}$ and $(\gt_n)_{1}^{\iy}$.
 It is possible to ``exclude'' from the spectral data
not the first eigenvalue $\t_1$ but an arbitrary norming constant
$\gt_m$, see \cite{KC09}.  Thus we define the spectral data
\[
\lb{J2} \gJ^1(2)=\Big\{ t\!=\!(t_n)_2^\iy: t_2<t_3<....., \qq
t_n\!=\!\t_n^o\!+\ve_{n\,},\qqq \ve=(\ve_{n-1\,})_1^\iy\!\in\!\ell^2
\Big\}.
\]
We recall known results about a RAB (i.e., a real-analytic
bijection) for inverse problems on a unit interval. We formulate
only results needed below.

\begin{theorem}
\lb{TMO} i)  The mapping $q\mapsto (\t \star \m)(q)$ defined by
\er{mntz1}, \er{st} is a RAB  between $\cL$ and $\gJ$.

\no ii)  The mapping $q\mapsto (\m\ts h_{(s)})(q)$ defined by
\er{mntz1}, \er{mntz2} is a RAB between $\cL$ and $\gJ^o\ts
\ell_1^2$.

\no iii) The mapping $(\t_n)_2^\iy\ts \gt$  from $\cL$ into
$\gJ^1(2)\ts \ell_1^2$   is a RAB between $\cL$ and $\gJ^1(2)\ts
\ell_1^2$.

\no iv) The mappings $ q\to (\n\ts \gh_{(s)})(q) $ from $\cL$ into
$\gJ^1\ts\el2_1$  is a RAB between $\cL$ and $\gJ^1\ts\el2_1$.
Moreover, all eigenvalues $(\n_n)_0^\iy$ and norming constants
$(\gh_{s,n})_0^\iy$ satisfy
\[
\lb{i1t}  {e^{\pm
\gh_{s,0}}\/|\dot\vt'(1,\n_0)|}-1=\sum_{n=1}^{+\iy} \rt(2-{e^{\pm
\gh_{s,n}}\/|\dot\vt'(1,\n_n)|}\rt),
\]
where the series converges absolutely and uniformly on bounded
subsets of $\cL$.

\end{theorem}

\no {\bf Remark.}\!  A bijection i) was proved in \cite{MO75}, see
\cite{PT87}, \cite{K19} about a RAB. The results of ii) were proved
in \cite{PT87}. The results of iii) were proved in \cite{KC09}. The
results of iv) were proved in \cite{KC09}, the proof is based on
\cite{IT83}  and the identities \er{i1t}.

In order to study $\cU_\s$, we describe basic properties of the
4-spectra mapping $\gf: \cL\to \ell^2$.

\begin{theorem}
\lb{Tf} i)   The 4-spectra mapping $\gf: \cL\to \ell^2$    defined
by
\[
\lb{gf}
 q\to \gf(q)=(\gf_n(q))_1^{\iy },\qq \gf_{2n-1}=\vr_n
-\t_n,\qqq  \gf_{2n}=\n_n -\m_n, \ \ \ \ \ n\ge 1,
\]
is a RAB between $\cL$ and $\ell^2$. Furthermore, the following
estimates hold true:
\[
\lb{esqf}
 \|q\|\le 2\|\gf\|(1+2\|\gf\|^{1\/3}),\ \ \ \
\|\gf\|\le 2\|q\|(1+2\|q\|^{1\/3}),
\]
where  $\|\gf\|^2=\sum_{n\ge 1}(|\n_n-\m_n|^2+|\t_n-\vr_n|^2)$.

\no ii) Let $q, q'\in \cL$. Then  the following identities hold
true:
\[
\lb{gf1}
\begin{aligned}
 \textstyle
   {q(0)-q(1)\/2}=\sum_{n\ge1} \gf_{2n-1},
   \qqq {q(0)+q(1)\/2}=\n_0+\sum_{n\ge1} \gf_{2n},
\\
   q(0)=\n_0+\sum_{n\ge1} \gf_{n},  \qqq
   q(1)=\n_0+\sum_{n\ge1}(-1)^n\gf_{2n}.
 \end{aligned}
\]
\end{theorem}

\no {\bf Proof.}  The results of i) were proved in \cite{K19}.
Identities \er{gf1} follow from \er{tr8}, \er{tr7} and  \er{tr6}.
\BBox

We consider the operator $ T y=-y''+qy$ on $ L^2(0,2)$ with
2-periodic conditions $y(x+2)=y(x), x\in \R$, where the potential
$q$ is 1-periodic and belongs to the real space $\cH$ defined by
$$
\cH=\Big\{f\in L^2(\T, \R): \int_0^1 fdx=0\Big\}, \qqq \T=\R/\Z.
$$
The spectrum of $ T $ is eigenvalues $\l^+_{0}, \l^\pm_{n}, n\ge 1$
which satisfy
$$
\begin{aligned}
& \l^+_{0}<\l^-_{1}\le \l^+_{1}<.... \le \l^+_{n-1}<
\l^-_n\le\l^+_{n}<...,
\\
& \l^\pm_{n}=(\pi n)^2+o(1)\qqq  \as \qq n\to \iy.
\end{aligned}
$$
These eigenvalues have the known  relations (see Fig. \ref{fig})
\[
\lb{exx}
\begin{aligned}
\n_0\le \l^+_0, \qqq \t_n, \vr_n\in (\l^+_{n-1},\l_{n}^-),\qqq  {\rm
and }\ \  \m_n, \n_n\in [\l^-_n,\l^+_{n}],\qqq \forall \ n\ge 1.
\end{aligned}
\]
  Here the equality $\l_n^-=\l_n^+$ means that $\l_n^-$ is a double
eigenvalue. The lowest eigenvalue $\l_0^+$ is simple, and the
corresponding eigenfunction has period 1. The eigenfunctions
corresponding to $\l_{n}^{\pm}$ have period 1 when $n$ is even and
they are antiperiodic, $y(x+1)=-y(x),\ x\in\R $, when $n$ is odd.

\setlength{\unitlength}{1.0mm}
\begin{figure}[h]
\centering
\unitlength 1.0mm 
\begin{picture}(135,25)
\put(5,10){\line(1,0){120.00}}

\put(15,10){\circle*{1}} \put(14,12.5){$\n_0$}

\put(28,6){$\l_0^+$} \put(30,10){\line(0,1){1.6}}
\put(30,11.6){\line(1,0){20.00}} \put(50,10){\line(0,1){1.6}}
\put(48,6){$\l_1^-$}

\put(35,9){$\ts$} \put(42,9){$\ts$} \put(35,13){$\t_1$}
\put(42,13){$\vr_1$}

\put(55,9){$\ts$} \put(61,9){$\ts$} \put(55,12.5){$\m_1$}
\put(61,12.5){$\n_1$}

\put(66,6){$\l_1^+$} \put(68,10){\line(0,1){1.6}}
\put(68,11.6){\line(1,0){22.00}} \put(90,10){\line(0,1){1.6}}
\put(88,6){$\l_2^-$} \put(74,9){$\ts$} \put(81,9){$\ts$}
\put(74,13){$\vr_2$} \put(81,13){$\t_2$}

\put(95,9){$\ts$} \put(102,9){$\ts$} \put(95,12.5){$\n_2$}
\put(102,12.5){$\m_2$}

\put(108,6){$\l_2^+$} \put(110,10){\line(0,1){1.6}}
\put(110,11.6){\line(1,0){15.00}}

\end{picture}

\caption{\footnotesize \!\!\! Periodic $\l_n^\pm$, Dirichlet $\m_n$,
Neumann $\n_n$ and mixed $\t_n$, $\vr_n$ eigenvalues.} \label{fig}
\end{figure}


For each $q\in \cL_0$ we consider the auxiliary Sturm-Liouville
problems on the interval $[0,2]$ with the Dirichlet and Neumann
boundary conditions with even potentials $\wt q$ on $[0,2]$:
\[
\lb{dqp} -f''+\wt q f=\l f,
\begin{aligned}
\qqq f(0)=f(2)=0,
\\
\qqq f'(0)=f'(2)=0,
\end{aligned}
\qqq
\wt q(x)=\ca q(x), &  0<x<1\\
                  q(2-x), & 1<x<2 \ac .
\]
 Let $\wt \m_n$ and $\wt \n_0, \wt \n_n, n\ge 1$ be
eigenvalues of the Dirichlet and Neumann problems respectively.

For a potential $\wt  q\in L^2(0,2)$ given by \er{dqp} we denote by
the same letter $\wt q$ its 2-periodic extension to the real line.
Introduce a operator $ \wt T y=-y''+\wt q y$ on $ L^2(0,4)$ with
4-periodic conditions, that is $y(x+4)=y(x), x\in \R$.
 The spectrum of $\wt T $ is a union of the eigenvalues $
\wt \l^+_{0}, \wt\l^\pm_{n}, n\ge 1$ which satisfy
$$
\begin{aligned}
& \wt \l^+_{0}<\wt \l^+_{1}\le \wt \l^+_{1}<.... \le \wt
\l^+_{n-1}<\wt \l^-_n\le\wt \l^+_{n}<...
\\
\textstyle &  \textstyle\wt \l^\pm_{n}=({\pi n\/2})^2+o(1)\qq \as
\qq n\to \iy.
\end{aligned}
$$
 Here the equality $\wt \l_n^-=\wt \l_n^+$
means that $\wt \l_n^-$ is a double eigenvalue. The lowest
eigenvalue $\wt \l_0^+$ is simple, and the corresponding
eigenfunction has period 2. The eigenfunctions corresponding to $\wt
\l_{n}^{\pm}$ have period 2 when $n$ is even and they are
antiperiodic, $y(x+2)=-y(x),\ x\in\R $, when $n$ is odd. It is well
known  the Dirichlet and Neumann  eigenvalues for even potentials
$\wt q$ satisfy
\[
\lb{gnc} \ti\n_0=\wt \l_0^+,\qqq \{\wt \l^-_n,\wt \l^+_{n}\}=\{\ti
\m_n,\ti\n_n\}\qq {\rm for \ all}\
 n\ge 1,
\]
see e.g., \cite{GT87}, \cite{KK97}. Recall standards results, see
e.g., \cite{K19}.

\begin{lemma} \lb{Tqp}
Let $\wt q$  be given by \er{dqp} for some $q\in \cL_0$. Then
Dirichlet $\wt\m_{n}$ and Neumann eigenvalues $\wt\n_{n-1} $ and
periodic eigenvalues $ \wt\l_{n}^\pm, n\ge 1$ satisfy
\[
\lb{iqp1}  \wt\m_{2n-1}=\t_{n}, \qqq  \wt\m_{2n}=\m_{n},
\]
\[
\lb{iqp2}   \wt\n_{2n-1}=\vr_{n},\qqq \wt\n_{2n}=\n_{n},
\]
\[
\lb{iqp3} \{\wt\l_{2n-1}^-, \wt\l_{2n-1}^+\} =\{\vr_{n},\t_{n}\},
\qqq \{\wt\l_{2n}^-,\wt\l_{2n}^+\} =\{\m_{n},\n_{n}\},
\]
where $\{A,B\}$ is a set of two elements $A,B$.
\end{lemma}

 The functions $\vp(1,\l), \vp'(1,\l)$ and $\vt(1,\l)$ are entire and
have the Hadamard factorizations
\[
\lb{E1}
\begin{aligned}
&    \vp(1,\l)=\prod_{1}^\iy{\m_n-\l\/ \m_n^o},\qq
\vp'(1,\l)=\prod_{1}^\iy{\t_n-\l\/ \t_n^o},
\\
&   \vt(1,\l)=\prod_{1}^\iy{\vr_n-\l\/ \vr_n^o}, \qq
   \vt'(1,\l)=(\n_0-\l)\vt_*(\l),\qqq  \vt_*(\l)=\prod_{1}^\iy
 {\n_n-\l\/ \n_n^o},
\end{aligned}
\]
uniformly in every bounded disk, see e.g., \cite{PT87}, \cite{M86}.

\subsection{Mappings $\cU_\s$ }
Now we describe properties of the mapping $\cU_\s$.

 \begin{proposition}
\lb{Tfs}

Let $\cU_\s=\gf^{-1}\s\gf: \cL\to \cL$ for some operator
$\s=(\s_j)_1^\iy\in \gS$. Then

\no i) The   4-periodic   eigenvalues
$\{\wt\l_0^+, \wt\l_{n}^\pm,n\ge 1\}$ are invariant under $\cU_\s$
and
\[
\lb{11}
(\wt\l_0^+,(\wt\l_{n}^\pm)_1^\iy)=(\wt\l_0^+,(\wt\l_{n}^\pm)_1^\iy)
\circ \cU_\s.
\]
 \no ii)  If  $ q^\bu=\cU_\s(q)$ for some $q\in \cL$, then for each
 $j\in \N$ we have
\[
\lb{U1}
{\rm if} \ n=2j-1 \Rightarrow \ca \t_j(q)=\vr_j(q^\bu),\qq
\vr_j(q)=\t_j(q^\bu) \ \ {\rm if}\qq  \s_n=-1
\\
\t_j(q)=\t_j(q^\bu),\qq \vr_j(q)=\vr_j(q^\bu) \ \ {\rm if} \qq
\s_n=1 \ac,
\]
\[
\lb{U2} {\rm if} \ n=2j \Rightarrow \ca \m_j(q)=\n_j(q^\bu),\qq
\n_j(q)=\m_j(q^\bu) \ \ {\rm if}\qq  \s_n=-1
\\
              \m_j(q)=\m_j(q^\bu),\qq \n_j(q)=\n_j(q^\bu)
              \ \ {\rm if} \qq \s_n=1 \ac .
\]
\no iii)  Let $\s_n=-1$ (or $\s_n=1$) for all odd $n\ge 1$. Let
$\s_n\in \{\pm
 1\}$ for all even  $n\ge 1$. Then   2-periodic   eigenvalues $\{\l_0^+,
\l_{n}^\pm,n\ge 1\}$ are invariant under $\cU_\s$ and
\[
\lb{11i} (\l_0^+,(\l_{n}^\pm)_1^\iy)=(\l_0^+,(\l_{n}^\pm)_1^\iy)
\circ \cU_\s.
\]
\no iv) Neumann  $\n_0$ is invariant under the mapping $\cU_\s$,
i.e.,  $\n_0=\n_0\circ \cU_\s$ for all $\s\in\gS$.
\end{proposition}

\no {\bf Proof.} The statements  i-ii) follow from Lemma
\ref{Tqp}.

\no iii) Consider the case when  $\s_n=-1$  for all odd $n\ge 1$,
the proof for $\s_n=1$ is similar.  Let $\s_n\in \{\pm
 1\}$ for all even  $n\ge 1$. Then from \er{U1} we have
$$
{\rm if} \ n=2j-1 \Rightarrow  \t_j(q)=\vr_j(q^\bu),\qq
\vr_j(q)=\t_j(q^\bu), \qq q^\bu=\cU_\s.
$$
These identities and \er{E1} imply
$$
2\D(\cdot,q)=\vp'(1,\cdot,q)+\vt(1,\cdot,q)
=\vt(1,\cdot,q^\bu)+\vp'(1,\cdot,q^\bu)=2\D(\cdot,q^\bu),
$$
which yields $\l_0^+(q)=\l_0^+(q^\bu), \l_n^\pm(q)=\l_n^\pm(q^\bu)$
for all $n\ge 1$.

iv) Let  $ q^\bu=\cU_\s(q)$ for some $q\in \cL$. Rewrite a Wronskian
in the form
$$f(\l,q)-g(\l,q)=1,\qq \where \qq
f(\l,q)=\vt(1,\l,q)\vp'(1,\l,q),\qq g(\l,q)=\vt'(1,\l,q)\vp(1,\l,q).
$$
The function $f(\cdot,q)$ is entire and has zeros $\vr_n, \t_n$ and
due to \er{U1} we obtain $f(\cdot,q)=f(\cdot,q^\bu)$ and then
$g(\cdot,q)=g(\cdot,q^\bu)$. The function $g(\cdot,q)$ is entire and
has zeros $\n_0(q)$ and a collection $A(q)=\{\n_n(q), \l_n(q),n\ge
1\} $. Note that due to \er{U2} we obtain $A(q)=A(q^\bu)$, which
yields   $\n_0(q^\bu)=\n_0(q)$. \BBox

We discuss properties of a mapping $U_1=\cU_\s=\gf^{-1}\s\gf$, where
$\s_n=-1$ for all $n\in \N$.

 \begin{lemma}
\lb{Tfx}

Let $q^{(1)}=U_1(q)$  for some $q\in \cL$ and $\wt q$ is given by
\er{dqp}. Then the eigenvalues for $q, q^{(1)}, \wt q$ satisfy
\[
\lb{12x} \ca \wt \m_{2n}=\m_{n}=\n_{n}(q^{(1)}),
           \\
\wt \m_{2n-1}=\t_{n}=\vr_{n}^{(1)} \ac,
           \qqq
            \ca \wt \n_{2n}=\n_{n}=\m_{n}^{(1)}
           \\  \wt \n_{2n-1}=\vr_{n}=\t_{n}^{(1)} \ac,
\]
\[
\lb{m3} \ca \{\wt \l_{2n-1}^-, \wt \l_{2n-1}^+\}
=\{\vr_{n},\t_{n}\}=\{\vr_{n}^{(1)},\t_{n}^{(1)}\}, \
\\
\{\wt \l_{2n}^-,\wt \l_{2n}^+\}
=\{\m_{n},\n_{n}\}=\{\m_{n}^{(1)},\n_{n}^{(1)}\}\ac,
\]
\[
\lb{m4} (\m, \n,\t,\vr)(q)=(\n,\m,\vr,\t)(q^{(1)}),
\]
for all $n\ge 1$, where $\m_{n}=\m_{n}(q),
\m_{n}^{(1)}=\m_{n}(q^{(1)}),...$ and $\wt \m_{n}=\m_{n}(\wt q),
...$ for shortness, and
\[
\lb{32} \vp'(1,\cdot,q)=\vt(1,\cdot,q^{(1)}), \qq
\vt(1,\cdot,q)=\vp'(1,\cdot,q^{(1)}),
\]
\[
\lb{32z} \textstyle \gh_{s,n}(q)=\ln |\vt(1,\n_n,q)|=\ln
|\vp'(1,\m_n^{(1)},q^{(1)})|=h_{s,n}(q^{(1)}).
\]
 \end{lemma}

\no {\bf Proof.} Fix $q\in \cL$ and the corresponding
$\gf(q)=(\gf_{n}(q))_1^\iy$. Define a mapping $q\to
\gf^\bu(q)=-\gf(q)\in \ell^2$. For this case there exists
$q^{(1)}\in \cL$  such that $\gf(q^{(1)})=\gf^\bu=-\gf(q)$, since
$\gf$ is a homeomorphism. Then we obtain
\[
\lb{11y} \ca -\gf_{2n-1}(q)=-\t_{n}(q)+\vr_{n}(q)=
 \gf_{2n-1}^\bu(q^{(1)}) =\t_{n}(q^{(1)})-\vr_{n}(q^{(1)}),
\\
-\gf_{2n}(q)=-\m_{n}(q)+\n_{n}(q)= \gf_{2n}^\bu(q^{(1)}) =
\m_{n}(q^{(1)})-\n_{n}(q^{(1)}), \ac \qq n\ge 1.
\]
Due to Lemma \ref{Tqp} we have
\[
\lb{m3z} \ca \{\wt \l_{2n-1}^-, \wt \l_{2n-1}^+\}
=\{\vr_{n},\t_{n}\}=\{\vr_{n}^{(1)},\t_{n}^{(1)}\} \
\\
\{\wt \l_{2n}^-,\wt \l_{2n}^+\}
=\{\m_{n},\n_{n}\}=\{\m_{n}^{(1)},\n_{n}^{(1)}\}\ac,
\]
and jointly with \er{iqp1}, \er{iqp2}, \er{11y} we obtain
\[
\lb{12xz}   \ca \wt \m_{2n}=\m_{n}=\n_{n}^{(1)},
           \\
\wt \m_{2n-1}=\t_{n}=\vr_{n}^{(1)} \ac,
           \qqq
            \ca \wt \n_{2n}=\n_{n}=\m_{n}^{(1)}
           \\
\wt \n_{2n-1}=\vr_{n}=\t_{n}^{(1)} \ac,
\qqq \ \forall \ n\ge 1.
\]
 These identities give  \er{12x}--\er{m4}.
Due to \er{gnc} the eigenvalues $\wt \n_{n}, \wt\m_{n}$ belong to
the ends of "gaps" $[\wt \l_{n}^-,\wt \l_{n}^+]$ for the even
potential $\wt q$. From \er{E1}, \er{m4} we have \er{32}, which
yields   \er{32z}. \BBox

We discuss properties of a mapping $U_o=\cU_\s$, where $\s_{j}=
(-1)^j$ for all $ j\in \N$.

 \begin{lemma}
\lb{Tfo}
 Let  $q\in \cL$ and $\wt q$ be given by \er{dqp}.
 Then eigenvalues for $q,
q^{(o)}=U_o(q)$ and $\wt q$ satisfy
\[
\lb{12xo} \ca \m_{n}=\m_{n}^{(o)}=\wt \m_{2n},
           \\
\t_{n}=\vr_{n}^{(o)}=\wt \m_{2n-1} \ac,
           \qqq
            \ca \n_{n}=\n_{n}^{(o)}=\wt \n_{2n}
           \\  \vr_{n}=\t_{n}^{(o)}=\wt \n_{2n-1} \ac,
\]
\[
\lb{m3o} \ca \{\wt \l_{2n-1}^-, \wt \l_{2n-1}^+\}
=\{\vr_{n},\t_{n}\}=\{\vr_{n}^{(o)},\t_{n}^{(o)}\},
\\
\{\wt \l_{2n}^-,\wt \l_{2n}^+\}
=\{\m_{n},\n_{n}\}=\{\m_{n}^{(o)},\n_{n}^{(o)}\}\ac,
\]
\[
\lb{m4o} (\m, \n,\t,\vr)(q)=(\m,\n,\vr,\t)(q^{(o)}),
\]
for all $n\ge 1$, where $\m_{n}=\m_{n}(q),
\m_{n}^{(o)}=\m_{n}(q^{(o)}),...$ and $\wt \m_{n}=\m_{n}(\wt q), ...$
for shortness, and
\[
\lb{32o} \vp'(1,\cdot,q)=\vt(1,\cdot,q^{(o)}), \qq
\vt(1,\cdot,q)=\vp'(1,\cdot,q^{(o)}),
\]
\[
\lb{32zo} \textstyle \gh_{s,n}(q)=\ln |\vt(1,\n_n,q)|=\ln
|\vp'(1,\m_n^{(o)},q^{(o)})|=h_{s,n}(q^{(o)}).
\]

 \end{lemma}

\no {\bf Proof.} Fix $q\in \cL$ and the corresponding
$\gf(q)=(\gf_{n}(q))_1^\iy$. Define a mapping
$$
q\to \gf^{(o)}(q)=((-1)^n\gf_n(q))_1^\iy\in \ell^2.
$$
 For this case there
exists $q^{(o)}\in \cL$  such that $\gf(q^{(o)})=\gf^\bu(q^{(o)})$,
since $\gf$ is a homeomorphism. Then we obtain
\[
\lb{11yo} \ca -\gf_{2n-1}(q)=-\t_{n}(q)+\vr_{n}(q)=
 \gf_{2n-1}^\bu(q^{(o)}) =\t_{n}(q^{(o)})-\vr_{n}(q^{(o)}),
\\
\gf_{2n}(q)=\m_{n}(q)-\n_{n}(q)= \gf_{2n}^\bu(q^{(o)}) =
\m_{n}(q^{(o)})-\n_{n}(q^{(o)}) \ac \qq n\ge 1.
\]
Due to Lemma \ref{Tqp} we have \er{m3o}
\[
\lb{m3zo} \ca \{\wt \l_{2n-1}^-, \wt \l_{2n-1}^+\}
=\{\vr_{n},\t_{n}\}=\{\vr_{n}^{(o)},\t_{n}^{(o)}\}, \
\\
\{\wt \l_{2n}^-,\wt \l_{2n}^+\}
=\{\m_{n},\n_{n}\}=\{\m_{n}^{(o)},\n_{n}^{(o)}\}\ac,
\]
and jointly with \er{iqp1}, \er{iqp2}, \er{11y} we obtain \er{12xo}.
 These identities give  \er{m3o}--\er{m4o}.
Due to \er{gnc} the eigenvalues $\wt \n_{n}, \wt\m_{n}$ belong to
the ends of gaps $(\wt \l_{n}^-,\wt \l_{n}^+)$ for the even
potential $\wt q$. From \er{E1}, \er{m4} we have \er{32o} and then
\er{32zo}. \BBox

\no {\bf Proof of Proposition \ref{TPr}} Due to the estimate
\er{esqf} the mapping $\cU_\s=\gf^{-1} \s\gf: \cL\to \cL$ is bounded
in any ball $\{\|q\|\le R\}$. The definition $\cU_\s=\gf^{-1} \s\gf:
\cL\to \cL$ implies that $\cU_\s=\cU_\s^{-1}$. The definition
$\cU_\s=\gf^{-1} \s\gf$ implies $\cU_\s\
\cU_{\s'}=\cU_{\s\s'}=\cU_{\s'}\ \cU_{\s}$ for all $\s, \s'\in \gS.$

Due to \er{11} the   4-periodic   eigenvalues $\{\wt\l_0^+,
\wt\l_{n}^\pm,n\ge 1\}$ are invariant under $\cU_\s$ and then the
Lyapunov function for the potential $\wt q(x)$ given by \er{dqp} is
also invariant under $\cU_\s$. Thus a functional  $\int_0^2|\wt
q(x)|^2dx$ is invariant under $\cU_\s$ (see e.g., \cite{K06}) and we
obtain for $r=\cU_\s (q)$:
$$
2\int_0^1|r(x)|^2dx=\int_0^2|(\wt r)(x)|^2dx=\int_0^2|\wt
q(x)|^2dx=2\int_0^1| q(x)|^2dx,
$$
which yields $\|\cU_\s(q)\|=\|q\|$.  \BBox


\section {Proof of main results for the unit interval}
\setcounter{equation}{0}

We discuss main results about 2-spectra mappings.

\no {\bf Proof of Theorem  \ref{T1}.} \no i)  From \er{m4} we obtain
$(\t\star \m)(q)=(\vr \star\n)(q^{(1)})$, where $q^{(1)}=U_1q$,
which yields $\t\star \m=(\vr\star \n)\circ U_1$. By Theorem
\ref{TMO}, the mapping $q\mapsto (\t\star \m)(q)$ from $\cL$ to
$\gJ$ is a RAB between $\cL$ and $\gJ$. Then  the mapping $q\mapsto
(\vr\star \n)(q)$ from $\cL$ to $\gJ$ is a RAB between $\cL$ and
$\gJ$, since due to Proposition \ref{TPr} $U_1$ is a RAB of $\cL$
onto itself.

 From \er{m4o} we obtain $(\t\star \m)(q)=(\vr
\star\m)(q^{(o)})$, where $q^{(o)}=U_oq$. Then due to Theorem
\ref{TMO}
 the mapping $q\mapsto (\vr\star \m)(q)$ from $\cL$ to
$\gJ$ is a RAB between $\cL$ and $\gJ$, since due to Proposition
\ref{TPr} $U_o$ is  an unitary operator on $\cL$.

Above we have obtained $\t\star \m=(\vr\star \n)\circ U_1$. From
\er{m4o} we have $\vr\star \n=(\t \star\n)\circ U_o$. A combination
of these identities gives $\t\star \m=(\t \star\n)\circ U_o\circ
U_1$. Then due to Theorem \ref{TMO}
 the mapping $q\mapsto (\vr\star \n)(q)$ from $\cL$ to
$\gJ$ is a RAB between $\cL$ and $\gJ$, since due to Proposition
\ref{TPr} $U_o, U_1$ are  RABs of $\cL$ onto itself. Collecting all
identities we obtain \er{enm}.

ii) Let $q\in \cL$ and let $\gn=\vr\star \n,\ \gn_{2j-1}=\vr_j,\
\gn_{2j}=\n_j$ be given. Above we have obtained $(\t\star
\m)(q)=(\vr\star \n)(q^{(1)})$, where $q^{(1)}= U_1 q$. This yields
$\t(q)=\vr(q^{(1)})$  and $\m(q)=\n(q^{(1)})$, which implies
$\t(q^{(1)})=\vr(q),\   \m(q^{(1)})=\n(q), $
 since $U_1^2=I$. Recall that  the recovering of a potential $q$ by \er{rq1}
via the 2-spectra mapping
 $\gn=\t\star \m,\ \gn_{2j-1}=\t_j,\ \gn_{2j}=\m_j, j\in \N$,
was obtained in \cite{K19}. Then we obtain the recovering a
potential $q^{(1)}$ by \er{rq1} via the 2-spectra mapping
$\gn=\vr\star \n,\ \gn_{2j-1}=\vr_j,\ \gn_{2j}=\n_j$. The proof of
iii) is similar.
 \BBox

Below we need identities, see e.g. \cite{PT87}: let $y$ be a
solution to the equation $-y''+qy=\l y$, then
\[
\lb{pti} \tes \int_0^1y^2(x,\l)dx=\{\dot y,y\}_w\big|_0^1, \qqq
\where\qqq \{f,y\}_w=fy'-f'y,\qq\dot y={\pa y\/\pa \l}.
\]
We discuss results about  mappings with eigenvalues plus norming (or
normolizing) constants.

 \no {\bf Proof of Theorem  \ref{T1x}.}
 i)   From \er{m4},  \er{32z} we obtain $\m\ts h_{(s)}=(\n\ts
\gh_{(s)})\circ U_1$. Then due to Theorem \ref{TMO}
 the mapping $q\mapsto \n \ts \gh_{(s)}$ from $\cL$ to
$\gJ^o\ts \ell_1^2$ is a  RAB between $\cL$ and $\gJ^o\ts \ell_1^2$,
since by Proposition \ref{TPr},  $U_1$ is a RAB of $\cL$ onto itself
and $\m\ts h_{(s)}:\cL\to \gJ^o\ts \ell_1^2$ is a RAB.

From \er{m4o}, \er{32zo} we obtain $\m\ts h_{(s)}=(\m \ts
\gh_{(s)})U_o$. Then due to Theorem \ref{TMO}
 the mapping $q\mapsto (\m\ts \gh_{(s)})(q)$ from $\cL$ to
$\gJ^o\ts \ell_1^2$ is a RAB between $\cL$ and $\gJ^o\ts \ell_1^2$,
since due to Proposition \ref{TPr} $U_o$ is an   unitary operator on
$\cL$.

Above we have obtained $\m\ts h_{(s)}=(\n\ts \gh_{(s)})\circ U_1$.
From \er{m4o} we have $\n\ts \gh_{(s)}=(\n\ts h_{(s)})\circ U_o$. A
combination of these identities gives
$$
 \m\ts h_{(s)}=(\n\ts \gh_{(s)})\circ U_1=
 (\n\ts h_{(s)})\circ U_o\circ U_1.
$$
  Then due to Theorem \ref{TMO}
 the mapping $q\mapsto (\n\ts \gh_{(s)})(q)$ from $\cL$ to
$\gJ^o\ts \ell_1^2$ is a RAB between $\cL$ and $\gJ^o\ts \ell_1^2$,
since due to Proposition \ref{TPr} $U_o, U_1$  are  RABs of $\cL$
onto itself. Collecting all  identities we obtain \er{enm1}.

\no ii) Consider  the mapping $\m\ts \a$. The identity \er{pti}
implies
 \[
 \lb{a3}
D_n=\dot\vp(1,\m_n)\vp'(1,\m_n),\qqq \forall \ n\in \N.
\]
Thus using \er{a3}, $D_n(0)={1\/2\m_n^o}$ and
$(-1)^n\vp'(1,\m_n)=e^{h_{s,n}}$ we rewrite $\a_n$ in the form
\[
\lb{a5}
 \begin{aligned} \tes
e^{\a_n}={D_n(q)\/D_n(0)}=[(-1)^n2\m_n^o\dot\vp(1,\m_n)] \
[(-1)^n\vp'(1,\m_n)]=e^{M_n+h_{s,n}},
\end{aligned}
 \]
where $M_n\in \R$ is defined by $0<(-1)^n2\m_n^o\dot\vp(1,\m_n)=
e^{M_n}$  and due to \er{ck1} the sequence $(M_n)_1^\iy\in
 \ell_1^2$. Thus we obtain
\[
 \lb{a6}
\a_n=h_{s,n}+M_n,\qqq \forall \ \ n\in \N.
\]
Due to Theorem \ref{TMO} the mapping $q\mapsto \m\ts h_{(s)}$ is a
bijection between $\cL$ and  $\gJ^o\ts \ell_1^2$ and the mapping
$\m\ts h_{(s)}\to \m\ts \a$ is a bijection from $\gJ^o\ts \ell_1^2$
onto itself, since $\a_n, h_{s,n}$ satisfy \er{a6}. This gives that
the mapping $q\mapsto (\m\ts \a)(q)$ is a bijection  between $\cL$
and $\gJ^o\ts \ell_1^2$.

Consider  the mapping $\n\ts \b$. The identity \er{pti} implies
 \[
 \lb{n3}
N_n(q)=-\dot\vt'(1,\n_n,q)\vt(1,\n_n,q).
\]
At $q=0$ we have
 \[
 \lb{n4} \tes
N_n(0)={1\/2},\qqq \dot\vt'(1,\n_n^o,0)=-{(-1)^n\/2}   ,\qqq
\vt(1,\n_n^o,0)=(-1)^n.
\]
 Then due to the definition \er{n2} of $\b_n$, and  \er{n3},\er{n4} and
 $(-1)^n\vt(1,\n_n,q)=e^{\gh_{s,n}}$  we get
\[
 \lb{n5}
\begin{aligned}
\tes
e^{\b_n}=2N_n=-2\dot\vt'(1,\n_n)(-1)^ne^{\gh_{s,n}}
=b_n[2(-1)^n\n_n^o\vt_*(\n_n)]e^{\gh_{s,n}}
=e^{\gh_{s,n}+K_n+K_n^o},
\end{aligned}
\]
where $K_n^o, K_n \in \R$ are defined by
$b_n={\n_n-\n_0\/\n_n^o}=e^{K_n^o}$ and  $0<(-1)^{n+1} 2\n_n^o
\dot\vt_*(\n_n)=e^{K_n}$ and we have used the Hadamard factorization
of $\vt'(1,\l)=-(\l-\n_0)\vt_*(\l)$ from \er{E1}. Thus we obtain
\[
 \lb{n7}
 \tes
\b_n=\gh_{s,n}+K_n+K_n^o, \qq K_n^o=\ln
{\n_n-\n_0\/\n_n^o}={O(1)\/n^2},
\]
and \er{ck3} gives $(K_n)_1^\iy\in \ell^2_1$. Repeating arguments
from the proof for $\a$  we obtain that the mapping $q\mapsto (\m\ts
\b)(q)$ is a bijection  between $\cL$ and $\gJ^o\ts \ell_1^2$.

We show the trace formulas \er{trN1}. The identity $\er{n5}$ yields
$e^{-\gh_{s,n}}=|\dot\vt'(1,\n_n)|/N_n$  and substituting one into
\er{i1t} we obtain ${1\/N_0}-1=\sum_{1}^{\iy} \big(2-{1\/N_n}\big)$,
i.e., we have  \er{trN1}, where the series converges absolutely and
uniformly on bounded subsets of $\cL$.

\no iii)  Let $q^\bu=U_1(q)$ and let $\m_n=\m_n(q),\a_n=\a_n(q),...
$ and $\m_n^\bu=\m_n(q^\bu),\a_n^\bu=\a_n(q^\bu),... $. From
\er{12x}, \er{a5}, \er{n5} we obtain for all $n\in \N$:
$$
\begin{aligned}
& \m_n^\bu=\n_n(q),\qqq
e^{M_n^\bu}=(-1)^n2\m_n^o\dot\vp(1,\m_n^\bu,q^\bu)
=(-1)^n2\n_n^o\dot\vt_*(\n_n,q)=e^{K_n}.
\end{aligned}
$$
From \er{32z} we have  $h_{s,n}^\bu=\gh_{s,n}$. Then these
identities and \er{n7} imply
$$
\a_n^\bu=h_{s,n}^\bu+M_n^\bu=\gh_{s,n}+K_n=\b_n-K_n^o=\hat\b_n,
$$
which yields $(\m\ts \a)\circ U_1=(\n\ts \hat\b)$. This gives that
the mapping $q\mapsto (\n\ts \hat\b)(q)$ is a bijection  between
$\cL$ and $\gJ^o\ts \ell_1^2$, since all other mappings $\m\ts \a$
and $U_1$ are bijections.
 \BBox

 {\bf Remark.} In \cite{SS10} the inverse problem for the mapping $q\to
\m\ts (\wt D_n)_1^\iy$ is discussed, where  $\wt D_n$ is the
normalizing constant given by  $\wt D_n= \m_n D_n(q)$. But the
presentation and the proof is not entirely clear.

We discuss a new type of inverse problems.  Let the Dirichlet
mapping $q\to \m=(\m_n)_1^\iy$ be given  and replace some $\m_n$ by
the Neumann eigenvalues $\n_n$. Then we obtain a replacing mapping
$\gc$. For example, we have $\gc=(\m_1,\n_2, \n_3,\m_4, \m_5,...)$.
There is a question: it is a good 1-spectra mapping?  We discuss
{\it replacing} mappings.

\begin{corollary}
\lb{T2} {\bf (Replacing mappings.)} Let $\s=(\s_n)_1^\iy$, where
$\s_{n}\in \{\pm 1\}$. Define replacing mappings $q\to
\ga=(\ga_n)_1^\iy, q\to\gb=(\gb_n)_1^\iy$ and
$q\to\gc=(\gc_n)_1^\iy$ and their components by
\[
\lb{dd1}
\begin{aligned}
\ga_n=\ca \t_n,\  {\rm if} \ \s_{2n-1}=1 \\
         \vr_n,\  {\rm if} \ \s_{2n-1}=-1\ac\!\!\!\!\!\!,\ \
          \gb_n=\ca \m_n,\qq  {\rm if} \ \s_{2n}=1 \\
         \n_n,\ {\rm if} \ \s_{2n}=-1\ac\!\!\!\!\!\!,\ \
         \gc_n=\ca  h_{s,n},\   {\rm if} \ \s_{2n-1}=1 \\
         \gh_{s,n},\ {\rm if} \ \s_{2n-1}=-1\ac\!\!\!\!\!\!.
\end{aligned}
\]

 \no i) Then the mapping $\ga\star \gb:\cL \to \gJ$ is a RAB
between $\cL$ and $\gJ$ and satisfies
\[
\lb{rm1} (\ga\star \gb)\circ \cU_\s=\t\star \m=(\vr\star \n)\circ
U_1=(\vr\star \m)\circ U_o=(\t\star \n)\circ U_o\circ U_1,
\]
and  all 2-spectra mappings $\ga\star \gb$, $\vr\star \n$, $\t\star
\n$, $\vr\star \m$ and $\t\star \m$ acting from $\cL$ into $\gJ$ are
isomorphic.

\no ii) The mapping $\gb\ts \gc: \cL \mapsto \gJ^o\ts \ell_1^2$ is a
RAB between $\cL$ and $\gJ^o\ts \ell_1^2$  and satisfies
\[
\lb{rm2} (\gb\ts \gc)\circ \cU_\s=\m\ts h_{(s)}=(\n \ts
\gh_{(s)})\circ U_1=(\m \ts \gh_{(s)})\circ U_o=(\n \ts
h_{(s)})\circ U_o\circ U_1,
\]
and all these mappings   are isomorphic.
 \end{corollary}

 \no {\bf Proof.} i) Let $q^\s=\cU_\s(q)$, where $q\in \cL$.
From \er{U1} we deduce that

if $ \s_{2n-1}=1$ then the eigenvalues
$\ga_n(q^\s)=\t_n(q^\s)=\t_n(q)$,

if $ \s_{2n-1}=-1$ then the eigenvalues
$\ga_n(q^\s)=\vr_n(q^\s)=\t_n(q)$.

From \er{U2} we deduce that

if $ \s_{2n}=1$ then the eigenvalues
$\gb_n(q^\s)=\m_n(q^\s)=\m_n(q)$,

 if $ \s_{2n}=-1$ then the eigenvalues $\gb_n(q^\s)=\n_n(q^\s)=\m_n(q)$.

Thus all these identities give  $\ga\star \gb=(\t\star \m)\circ
\cU_\s$. Then due to Theorem \ref{TMO}
 the mapping $q\mapsto (\ga\star \gb)(q)$ from $\cL$ to
$\gJ$ is a RAB between $\cL$ and $\gJ$, since the mapping $U_\s:
\cL\to \cL$ is a RAB from $\cL$ onto itself. Moreover, we have
\er{rm1}.
 The proof of ii) is similar. \BBox

\section {Mixed boundary conditions}
\setcounter{equation}{0}

We discuss isomorphic inverse problems  for mixed eigenvalues.
Recall that we have introduced norming constants $\gt_n, \gr_n$
(associated with mixed eigenvalues $\t_n, \vr_n$ respectively) and
the corresponding mappings by
\[
\label{nce4q}
\begin{aligned}
\gt_n=\ln |\vp(1,\t_n)\sqrt{\t_n^o}|,& \qqq \gr_n(q)=-\ln
|\vt'(1,\vr_n)/\sqrt{\t_n^o}|=\ln |\vp(1,\vr_n)\sqrt{\vr_n^o}|,
\\
q\to \gt=(\gt_n)_1^\iy,  \qqq & \qqq q\to \gr=(\gr_n)_1^\iy,
\end{aligned}
\]
where $\gt(q)\in \ell_1^2$  if $q\in \cL$, see \cite{KC09}. We
consider the known facts about properties of Sturm-Liouville
problems under the reflection (unitary) operator $\cR: \cL\to \cL$
defined by $(\cR y)(x)=y(1-x), x\in (0,1)$. Let $y_n(x,q), n\ge 1$
be  the eigenfunction corresponding be the Dirichlet eigenvalue
$\m_n(q)$:
$$
-y_n''+qy_n=\m_n y_n,\qq q\in \cL.
$$
Then the function $u_n=\cR y_n$ satisfies $ -u_n''+q^\bu u_n=\m_n
u_n $, where $q^\bu=\cR q$ and then $\m_n(q)=\m_n(q^\bu)$. We apply
similar arguments for the Neumann and mixed eigenvalues and we
obtain
\[
\lb{R1} (\m_n, \n_n,\t_n,\vr_n)=(\m_n,\n_n,\vr_n,\t_n)\circ \cR\qq
\forall \ n\ge 1.
\]
In particular, it gives that the two mappings $\t\ts \gt$ and $\t\ts
\gr$ acting from $\cL$ into $\gJ^1(2)\ts \ell_1^2$  are unitarily
equivalent.  We discuss  the mappings $q\to \t^{(2)}=(\t_n)_2^\iy$
and  $q\to \vr^{(2)}=(\vr_n)_2^\iy$ and formulate results  based on
\cite{KC09} about inverse problems for mixed eigenvalues. In this
case we obtain unitarily equivalent mappings.

 \begin{proposition}
\lb{Tmix} i)  The operator $U_o=\cR$ and satisfies
\[
\lb{eenm} \vr\ts \gr=\big(\t\ts \gt\big) \circ U_o.
\]
The two mappings $\t\ts \gt$ and $\t\ts \gr$ acting from $\cL$ into
$\gJ^1(2)\ts \ell_1^2$  are unitarily equivalent.

\no ii)   Let $\gJ^1(2)$ be defined by \er{J2}. Define a mapping
$q\to \vr^{(2)}=(\vr_n)_2^\iy$. Then a mapping $\vr^{(2)}\ts \gr$ is
a RAB between $\cL$ and $\gJ^1(2)\ts \ell_1^2$. Moreover, the
following trace formulas hold true
\[
\label{b1x} \sum_{n=1}^{\iy} \lt(2-{e^{\gt_n}\/\sqrt {\t_n^o}\ |
{\vp'\/\pa \l}(1,\t_n)|}\rt)=0,\qqq
  \sum_{n=1}^{+\iy} \lt(2-{e^{\gr_n(q)}\/\sqrt {\t_n^o}
|{\pa \vt\/\pa \l}(1,\r_n,q)|}\rt)=0.
\]

\end{proposition}

\no {\bf Proof.} i) Due to \er{R1} the mappings $\vr$ and $\t$ from
$\cL$ into $\gJ^1$ are unitarily equivalent and satisfy $\vr=\t\circ
\cR$. From \er{R1}, \er{enm} we obtain $\t\star \m=(\vr\star
\m)\circ U_o=(\vr\star \m)\circ \cR$, which yields that $\cR=U_0$,
since due to Theorem \ref{T1} the mappings $\t\star \m $ and
$\vr\star \m$ are bijections.

From  $\cR=U_o$ and  \er{R1} we have $\m_{n}(q)=\m_{n}(q^\bu),
q^\bu=U_o q$, which jointly with \er{E1}   yield
$\vp(1,\l,q)=\vp(1,\l,q^\bu)$. Then the definitions \er{nce4q} and
\er{R1} imply
\[
\lb{Me33}
\begin{aligned}
\gt_n(q)=\ln |\vp(1,\t_n(q),q)\sqrt{\t_n^o}|=\ln
|\vp(1,\vr_n(q^\bu),q^\bu)\sqrt{\t_n^o}|=\gr_n(q^\bu),\qqq \forall \
n\ge 1,
\end{aligned}
\]
which jointly with  $\cR=U_0$,  and \er{R1} yield \er{eenm}.

ii) Due to Theorem \ref{TMO} and \er{eenm} the mapping $\vr^{(2)}\ts
\gr$ is a RAB between $\cL$ and $\gJ^1(2)\ts \ell_1^2$,  since
$U_o=\cR$ is an unitary operator on  $\cL$.

The first trace formula in \er{b1x} was proved in \cite{KC09}.    We
show the second one. Identities \er{32o} give
$\vp'(1,\l,q)=\vt(1,\l,q^\bu)$. Then substituting this identity,
$\t_{n}(q)=\vr_{n}(q^\bu)$ from \er{R1} and \er{Me33} into the first
trace formula in \er{b1x} we obtain the second one in \er{b1x}.
\BBox

 Consider inverse problems for a mapping $q\to \t\ts
(\cD_n)_1^\iy$, where $\cD_n$ is a normalizing constant defined by
\[
\lb{mi1} \tes \cD_n(q)=\int_0^1\vp^2(x,\t_n,q)dx,\qq n\in \N, \qqq
\where \qq \cD_n(0)={1\/2\t_n^o}.
\]
It is more convenient to modify constants $\cD_n$ and define another
mapping $q\to \e=(\e_n)_1^\iy$, where the components $\e_n$ are
given by
\[
\lb{mi2}\tes \e_n=\log {\cD_n(q)\/\cD_n(0)}=\log \big[2\t_n^o
\cD_n(q)\big], \qqq \qq n\in \N.
\]
Consider inverse problems for a mapping $q\to \vr\ts (\cN_n)_1^\iy$,
where $\cN_n$ is a normalizing constant defined by
\[
\lb{mi6} \tes
 \cN_n(q)=\int_0^1\vt^2(x,\vr_n,q)dx,\qq n\in \N, \qqq
\where \qq \cN_n(0)={1\/2}.
\]
It is more convenient to define modified normalizing constant $\c_n$
given
\[
\lb{mi7} \tes \c_n=-\log {\cN_n(q)\/\cN_n(0)}=-\log \big[2
\cN_n(q)\big], \qqq \qq n\in \N.
\]
and introduce a  mapping $q\to \c=(\c_n)_1^\iy$. We discuss results
about inverse problems for mixed boundary conditions.

\begin{theorem}
\label{Tmix2}
i)  Let $\s_{2n}=1$ and  $\s_{2n-1}\in \{\pm 1\}$ for all $ n\in
\N$. Define replacing mappings $q\to \ga=(\ga_n)_2^\iy$ and
$q\to\gc=(\gc_n)_1^\iy$, where the components are given by
\[
\lb{dd1m}
\begin{aligned}
\ga_n=\ca \t_n,\  {\rm if} \ \s_{2n-1}=1 \\
         \vr_n,\  {\rm if} \ \s_{2n-1}=-1\ac\!\!\!\!\!\!,\ \
                   \gc_n=\ca  \gt_{n},\   {\rm if} \ \s_{2n-1}=1 \\
         \gr_{n},\ {\rm if} \ \s_{2n-1}=-1\ac\!\!\!\!\!\!.
\end{aligned}
\]
Then the two mappings $\ga\ts \gc$ and $\t^{(2)}\ts \gt$ from $\cL
\to \gJ^1(2)\ts \ell_1^2$ are isomorphic. Moreover, each of them is
a RAB between $\cL$ and $\gJ^1(2)\ts \ell_1^2$ and satisfies
\[
\lb{mqq} \ga\ts \gc=(\t^{(2)}\ts \gt)\circ \cU_\s.
\]
ii) Each of two mappings $\t^{(2)}\ts \e$ and $\vr^{(2)}\ts \c$
acting from $\cL$ into $\gJ^1(2)\ts \ell_1^2$ is a bijection between
$\cL$ and $\gJ^1(2)\ts \ell_1^2$.
\end{theorem}

\no {\bf Remark.} 1) The mapping $q\to \t\ts (\cD_n)_1^\iy$ is
considered in \cite{SS08}, but the authors do not see that spectral
data $\t, (\cD_n)_1^\iy$ are dependent due to \er{b1x}.

\no{\bf Proof.} i)  Let $q^\s=\cU_\s(q)$. Consider $\ga$. From
\er{U1} we deduce that
 the eigenvalues
\[
\lb{xx1} \ga_n(q^\s)=\ca \t_n(q^\s)=\t_n(q) \ \ {\rm if} \
\s_{2n-1}=1
\\
                  \vr_n(q^\s)=\t_n(q)\  \ {\rm if} \  \s_{2n-1}=-1 \ac.
\]
This yields $\ga_n(q^\s)=\t_n(q)$ for all $n\ge 2$.

Consider the mapping $\gc$: if $ \s_{2n-1}=1$, then due to \er{E1},
\er{xx1} the norming constant
$$
\begin{aligned}
\gc_n(q^\s)=\gt_n(q^\s)=\ln|\vp(1,\t_n(q^\s),q^\s)\sqrt{\t_n^o}|=
\ln|\vp(1,\t_n(q^\s),q)\sqrt{\t_n^o}|
\\
= \ln|\vp(1,\t_n(q),q)\sqrt{\t_n^o}|=\gt_n(q),
\end{aligned}
$$
if $ \s_{2n-1}=-1$ then due to \er{E1}, \er{xx1} the norming
constant
$$
\begin{aligned}
\gc_n(q^\s)=\gr_n(q^\s)=\ln|\vp(1,\t_n(q^\s),q^\s)\sqrt{\t_n^o}|=
\ln|\vp(1,\vr_n(q),q)\sqrt{\t_n^o}| =\gt_n(q).
\end{aligned}
$$
 This yields
$\gc_n(q^\s)=\gt_n(q)$ for all $n\in \N$. Collecting the identities
we obtain \er{mqq}.
 Then due to Theorem \ref{TMO}
 the mapping $q\mapsto (\ga\ts \gc)(q)$ from $\cL$ to
$\gJ^1(2)\ts \ell_1^2$ is a RAB between $\cL$ and $\gJ^1(2)\ts
\ell_1^2$, since due to Proposition \ref{TPr} $U_\s$ is a RAB of
$\cL$ onto itself.

\no ii) Consider the mapping $\t^{(2)}\ts \e$. The identity \er{pti}
implies
\[
 \lb{mi3}
\cD_n=-\dot\vp'(1,\t_n)\vp(1,\t_n),\qqq \forall \ n\in \N.
\]
Then using  $(-1)^{n+1}\vp(1,\t_n)\sqrt{\t_n^o}=e^{\gt_n}$ we
rewrite $\e_n$ in the form
\[
 \lb{mi4}
 \begin{aligned} \tes
e^{\e_n}={\cD_n(q)\/\cD_n(0)}=[(-1)^n2\sqrt{\t_n^o} \
\dot\vp'(1,\t_n)] \
[(-1)^{n+1}\sqrt{\t_n^o}\vp(1,\t_n)]=e^{\cM_n+\gt_n},
\end{aligned}
 \]
where $\cM_n\in \R$ is defined by $0<(-1)^n2\sqrt{\t_n^o} \
\dot\vp'(1,\t_n)= e^{\cM_n}$ and due to \er{ck2} the sequence
$(\cM_n)_1^\iy\in \ell_1^2$. Thus we obtain
\[
 \lb{mi5}
\e_n=\gt_n+\cM_n,\qqq \forall \ \ n\in \N.
\]
The mapping $q\mapsto \t^{(2)}\ts \gt$ is a bijection between $\cL$
and $\gJ^1(2)\ts \ell_1^2$ and the mapping $\t^{(2)}\ts \gt\to
\t^{(2)}\ts \e$ is a bijection from $\gJ^1(2)\ts \ell_1^2$ onto
itself, since $\e_n, \gt_{n}$ satisfy \er{mi5}. This gives that the
mapping $q\mapsto (\t^{(2)}\ts \e)(q)$ is a bijection  between $\cL$
and $\gJ^1(2)\ts \ell_1^2$.

$\bu$ Consider the mapping $\t^{(2)}\ts \c$. The identity \er{pti}
implies
\[
 \lb{mi8}
\cN_n=\dot\vt(1,\vr_n)\vt'(1,\vr_n),\qqq \forall \ n\in \N.
\]
Then using   $e^{-\gr_n}=(-1)^{n+1}\vt'(1,\vr_n)/\sqrt{\vr_n^o}$ we
rewrite $\c_n$ in the form
\[
 \lb{mi4x}
 \begin{aligned}  \tes
e^{-\c_n}={\cN_n(q)\/\cN_n(0)}=[(-1)^{n}2\sqrt{\vr_n^o} \
\dot\vt(1,\vr_n)] \
[(-1)^{n}\vt'(1,\vr_n)/\sqrt{\vr_n^o}]=e^{-\cK_n-\gr_n},
\end{aligned}
 \]
where $\cK_n\in \R$ is defined by
$e^{-\cK_n}=(-1)^{n}2\sqrt{\vr_n^o} \ \dot\vt(1,\vr_n)>0 $ and due
to \er{ck3} the sequence $(\cK_n)_1^\iy\in \ell_1^2$. Thus we obtain
\[
 \lb{mi5q}
\c_n=\gr_n+\cK_n,\qqq \forall \ \ n\in \N.
\]
The mapping $q\mapsto \vr^{(2)}\ts \gr$ is a bijection between $\cL$
and $\gJ^1\ts \ell_1^2$ and the mapping $\vr^{(2)}\ts \gr\to
\vr^{(2)}\ts \e$ is a bijection from $\gJ^1(2)\ts \ell_1^2$ onto
itself, since $\e_n, \gr_{n}$ satisfy \er{mi5}. This gives that the
mapping $q\mapsto (\vr^{(2)}\ts \c)(q)$ is a bijection  between
$\cL$ and $\gJ^1(2)\ts \ell_1^2$. \BBox

\section {Smooth potentials  }
\setcounter{equation}{0}

\subsection{Smooth potentials}

Consider  the case of potentials from Sobolev spaces $\cL_k$ given
by
$$
\cL_k=\{q,  q^{(k)}\in \cL\},  \qq  k\ge 0, \qq \cL=\cL_0.
$$
Following the book of P\"oschel and Trubowitz \cite{PT87} in analogy
to the notation $O(1/n)$, we use the notation $\ell_k^2(n)$ for an
arbitrary sequence of numbers which is an element of $\ell_k^2$:
$$
y_n=y_n^o+\ell_k^2(n)  \Longleftrightarrow  \qqq \sum_{n\ge 1}n^{2k}
|y_n-y_n^o|^2<\iy.
$$

We define the spectral data $\gJ_k$ for potentials from $\cL_k, k\in
\N$ by
$$
\gJ_k=\Big\{(s_n)_1^\iy\in  \gJ: \sqrt{s_n}={\textstyle{\pi
n\/2}}+\sum_{1\le j\le d }{a_j\/(\pi n)^{2j+1}}
 +\ell_{k+1}^2(n), \qq (a_j)_1^d\in \R^d\Big\},\qq \tes d=[{k+1\/2}],
$$
here $[r]$ is the integer part of $r\in \R$ and  the coefficients
$a_j$ depend on a sequence $s=\!(s_n)_1^\iy$. We define two sets
$\gJ_k^o$ and $\gJ_k^1$ of all real, strictly increasing sequences
by
$$
\gJ_k^o=\Big\{ s\!=\!(s_n)_1^\iy\in \gJ^o: \sqrt{s_n}=\pi
n+\sum_{1\le j\le d}a_j\ve^{2j+1} +\ell_{k+1}^2(n), \qq (a_j)_1^d\in
\R^{d} \Big\},\ \ \tes \ve={1\/2\pi n},
$$
$$
\begin{aligned}
 \gJ_k^1=\Big\{\!(s_n)_1^\iy\in \gJ^1: \sqrt{s_n}=\pi
{\tes (n-{1\/2})}+\sum_{1\le j\le d}b_j \d_n^{2j+1}
 +\ell_{k+1}^2(n), \ \ (b_j)_1^d\in
\R^d \Big\},\ {\tes  \d_n={1\/2\pi (n-{1\/2})},}
\end{aligned}
$$
Note that the coefficients  $a_j, b_j$ depend on a sequence
$s=\!(s_n)_1^\iy$. Recall results from \cite{MO75}: if $q\in \cL_k$,
then we have asymptotics of the Dirichlet eigenvalues $ \m_{n}$:
\[
\lb{Dm1}
\begin{aligned}
& \sqrt{ \m_n}=\pi n+\sum_{1\le j\le d}{a_j\ve^{2j+1}} +\ve
^{k+1}\wt r_n,\qq  \wt r_n=q_\bu^{(k)}(n)+\ell_{1}^2(n),\qq (\wt
r_n)_1^\iy\in \ell^2,
\end{aligned}
\]
where the coefficients  $(a_j)_1^d\in \R^{d}$ depend on $q$, \  $
d=\big[{k+1\/2}\big],$ and $\wt r_n$ has the form
\[
\lb{fcs}
\begin{aligned}
q_\bu^{(k)}(n)=(-1)^{1+[{k\/2}]} \int_0^1 q^{(k)}(x)\cF_k(nx)dx, \qq
\cF_k(nx)=\ca \sin[2\pi nx], \qq k+1\in 2\N
\\  \cos[2\pi nx], \qq k\in 2\N \ac ,
\end{aligned}
\]
and asymptotics of the mixed eigenvalues $\t_{n}$:
\[
\lb{Dm2}
\begin{aligned}
\sqrt{ \t_n}= \pi n'+\sum_{1\le j\le d} \wt\t_j \d_n^{2j+1} + \wt
q_\bu^{(k)}(n)\d_n^{k+1}  +\ell_{k+2}^2(n)\qqq \as \qq n\to \iy,\qq
{\tes n'=n-{1\/2}},
\\
\wt q_\bu^{(k)}(n)=(-1)^{1+[{k\/2}]} \int_0^1
q^{(k)}(x)\cF_k(n'x)dx, \qq \cF_k(nx)=\ca \sin[2\pi n'x], \qq k+1\in
2\N
\\  \cos[2\pi n'x], \qq k\in 2\N \ac ,
\end{aligned}
\]
where the coefficients  $\wt\t_j\in \R,\  j\in \N_d$ depend on $q$.
We recall the famous results of Marchenko and Ostrovski (Corollary
5.1 from \cite{MO75}) about the mapping $\t\star \m$ from $\cL_k$ to
$\gJ_k$.

 \begin{theorem}
\lb{TB} Each 2-spectra mapping $q\mapsto (\t \star \m)(q)$ is a
bijection between $\cL_k$ and $\gJ_k, k\in\N$.
\end{theorem}

Asymptotics \er{Dm1}, \er{Dm2} give a 2-spectra mapping $q\to\t\star
\m$ from $\cL_k$ to $\gJ_k$ for any $k\in \N$. We show that the
mappings $\t\star \m$, $\vr\star \n$, $\t\star \n$ and $\vr\star \m$
are isomorphic bijections between $\cL_k$ and $\gJ_k$. Moreover, we
show that the mappings $\m\ts h_{(s)}, \n \ts \gh_{(s)}$,  $\m \ts
\gh_{(s)}, \n \ts h_{(s)}$ are isomorphic bijections between $\cL_k$
and $\gJ_k^o\ts \gJ_k^e$.

\begin{theorem}
\lb{Tsq1}

\no i) The mappings $U_1, U_o$ are bijections of $\cL_k$ onto itself
for any $k\in \N$.

\no ii) All 2-spectra mappings $\vr\star \n$, $\t\star \n$,
$\vr\star \m$ and $\t\star \m$ acting from $\cL_k$ into $\gJ_k$ are
isomorphic  for any $k\in \N$, and each of them is a bijection
between $\cL_k$ and $\gJ_k$ and satisfy \er{enm}.

\no iii)  Mappings $\m\ts h_{(s)}, \n \ts \gh_{(s)}$,  $\m \ts
\gh_{(s)}$ and $\n \ts h_{(s)}$ (defined by  \er{mntz1}, \er{mntz2})
acting from $\cL_k$ into $\gJ_k^o\ts \gJ_k^e$ are isomorphic for any
$k\in \N$, and each of them is a bijection between $\cL_k$ and
$\gJ_k^o\ts \gJ_k^e$ and satisfy
\[
\lb{enm11} \begin{aligned} \m\ts h_{(s)}=(\n \ts \gh_{(s)})\circ
U_1=(\m \ts \gh_{(s)})\circ U_o=(\n \ts h_{(s)})\circ U_o\circ U_1.
\end{aligned}
\]
 \end{theorem}

\no {\bf Proof.} i) Let  $q\in \cL_k$ for some $k\ge 1$. Repeating
arguments from \cite{MO75} for the case \er{Dm1}, \er{Dm2} we
determine  asymptotics the eigenvalues $(\n_n)_1^\iy$ and
$(\vr_n)_1^\iy$ of the boundary value problems $y'(0)=y'(1)=0$ and
$y'(0)=y(1)=0$:
\[
\lb{nm1}
\begin{aligned}
\sqrt{ \n_n}=\pi n+\sum_{1\le j\le d}\wt a_j \ve ^{2j+1}
+\ell_{k+1}^2(n),\qq \tes  \ve={1\/2\pi n},
\end{aligned}
\]
\[
\lb{nm2}
\begin{aligned}
\sqrt{ \vr_n}=\pi n' +\sum_{1\le j\le d}\wt b_j \d_n^{2j+1}
+\ell_{k+1}^2(n), \qq \tes  \d_n={1\/2\pi n'},\qq {\tes n'=n-{1\/2}}
\end{aligned}
\]
for some constants $(\wt a_j)_1^d, (\wt b_j)_1^d\in \R^{d}$
depending on $q$.

 Due to \er{enm} the Dirichlet $\m_{n}$,
Neumann $\n_{n}$ and mixed $\t_{n},\vr_{n}$   eigenvalues satisfy
\[
\lb{10x} (\m_{n},
\n_{n},\t_{n},\vr_{n})(q)=(\n_{n},\m_{n},\vr_{n},\t_{n})(q^\bu),
\qqq \forall \ n\ge1,\ \  \where \qq  q^\bu=U_1(q).
\]
 Then the Dirichlet $\m_{n}(q^\bu)$,  and mixed $\t_{n}(q^\bu)$
eigenvalues have the corresponding asymptotics \er{Dm1}, \er{Dm2}.
Thus from Theorem \ref{TB} we deduce that $q^\bu=U_1(q)\in \cL_k$,
i.e., $U_1 \cL_k\ss \cL_k$, and $U_1^2=I$ gives $U_1 \cL_k= \cL_k$.

\no ii) We will show that 2-spectra mappings $\vr\star \n$, $\t\star
\n$, $\vr\star \m$ and $\t\star \m$ acting from $\cL_k$ into $\gJ_k$
are isomorphic. Consider the 2-spectra mapping $\vr\star \n$ acting
from $\cL_k$ into $\gJ_k$. The proof for other mappings is similar.
In i) we have obtained that if $q\in \cL_k$, then the eigenvalues
$(\n_n)_1^\iy$ and $(\vr_n)_1^\iy$ alternate and the asymptotics
\er{nm1}, \er{nm2} hold true, this yields $q\to \vr\star \n$ is a
mapping from $\cL_k$ into $\gJ_k^\bu$.

In \er{10x} we have an identity   $\t\star \m=(\vr\star \n)\circ
U_1$, which yields $(\t\star \m)\circ U_1=\vr\star \n$. Then
  the mapping $\vr\star \n: \cL_k\to
\gJ_k$ is a bijection between $\cL_k$ and $\gJ_k$, since due to
Theorem \ref{TB} the mapping $\t\star \m: \cL_k\to \gJ_k$ is a
bijection between $\cL_k$ and $\gJ_k$ and $U_1: \cL_k\to \cL_k$ is a
bijection from $\cL_k$ onto itself. Moreover, due to the identity
$\t\star \m=(\vr\star \n)\circ U_1$ the mappings $\t\star \m$ and
$\vr\star \n$   are isomorphic.

\no iii)  Due to \er{10x}, \er{y} we have a mapping $q\mapsto (\m
\ts h_{(s)})(q)$ from $\cL_k$ into $\gJ_k^o\ts \gJ_k^e$ for any
integer $k\in \N$. Theorem \ref{T1x} gives an injection of this
mapping.

In order to  show a surjection of this mapping we use arguments from
the proof of Theorem \ref{TB}  from \cite{MO75}. Let $\m^\cd\ts
h_{(s)}^\cd\in \gJ_k^o\ts \gJ_k^e$ be given for some $k\ge 1$. Then
due to Theorem \ref{TMO}, ii) there exist unique $q\in \cL_0$ such
that $\m(q)\ts h_{(s)}(q)=\m^\cd\ts h_{(s)}^\cd$. Assume that $q\in
\cL_m$ but $q\notin \cL_{m+1}$, where $m<k$. Further we actually
repeat the proof from \cite{MO75} verbatim and using sharp
asymptotics \er{y} of $h_{s,n}$ and \er{Dm1} of $\m_n$ show that
$q\in \cL_{m+1}$, which gives a contradiction. Here it is important
to determine  the new sharp asymptotics of $h_{s,n}$ from \er{y}
with the Fourier coefficients \er{yc}.

Using the identities  \er{enm1}, and the bijection of the mapping
$q\to\m(q)\ts h_{(s)}(q)$ acting from $\cL_k$ into $\gJ_k^o\ts$ we
have the proof of iii), since by i),  the mappings $U_1, U_o$ are
bijections of $\cL_k$ onto itself for any $k\in \N$.
 \BBox

We discuss inverse problems for mixed eigenvalue mapping
$\t^{(2)}=(\t_n)_2^\iy$ for potentials $q\in \cL_k, k\in \N$. In
this case we define the corresponding spectral data $\gJ_k^1(2)$ by
\[
\lb{mJ2} \gJ_k^1(2)=\Big\{ s\!=\!(s_n)_2^\iy\in \gJ^1:
s_n=\sum_{1\le j\le d}{a_j\/ (2\pi n)^{2j+1}} +\ell_{k+1}^2(n), \qq
(a_j)_1^d\in \R^{d} \Big\},\qq k\in \N,
\]
and  a set of all possible norming constants $\gt=(\gt_{n})_1^\iy$
by
$$
\gJ_k^e=\Big\{ s\!=\!(s_n)_1^\iy\in \ell_1^2: s_n=\sum_{1\le j\le
d}{\wt a_j\/ (2\pi (n-{1\/2})^{2j}} +\ell_{k+1}^2(n), \qq (\wt
a_j)_1^d\in \R^{d} \Big\},\qq k\in \N,
$$
Note that   $(a_j)_1^d\in \R^{d}$ and  $(\wt a_j)_1^d\in \R^{d}$
depends on $q$.

\begin{theorem}
\lb{Tmb}
 Let $\t^{(2)}=(\t_n)_2^\iy$.
The mapping $\t^{(2)}\ts \gt$ from $\cL_k$ into $\gJ_k^1(2)\ts
\gJ_k^e$ is a bijection between $\cL_k$ and $\gJ_k^1(2)\ts \gJ_k^e$.
 \end{theorem}

\no {\bf Proof.} Due to \er{Dm2}, \er{my} we have a mapping
$q\mapsto (\t^{(2)}\ts \gt)(q)$ from $\cL_k$ into $\gJ_k^1(2)\ts
\gJ_k^e$ for any integer $k\in \N$. Theorem \ref{TMO} iii) gives an
injection of this mapping.

In order to  show a surjection of this mapping we use arguments from
the proof of Theorem \ref{TB}  from \cite{MO75}. Let $\t_\cd\ts
\gt_\cd\in \gJ_k^1(2)\ts \gJ_k^e$  be given for some $k\ge 1$. Then
due to Theorem \ref{TMO}, ii) there exist unique $q\in \cL_0$ such
that $(\t^{(2)}\ts \gt)(q)=\t_\cd\ts \gt_\cd$. Assume that $q\in
\cL_m$ but $q\notin \cL_{m+1}$, where $m<k$. Further we actually
repeat the proof from \cite{MO75} verbatim and using sharp
asymptotics \er{Dm2} of $\t_n$ and \er{my} of $\gt_n$ show that
$q\in \cL_{m+1}$, which gives a contradiction. Here it is important
to have the sharp asymptotics of $\t_n, \gt_n$ from \er{Dm2},
\er{my} with the Fourier coefficients. \BBox

The identities \er{eenm}, i.e., $ \vr^{(2)}\ts \gr=\big(\t^{(2)}\ts
\gt\big) \circ U_o$, and the bijection of the mapping
$q\to(\t^{(2)}\ts \gt)(q)$ between $\cL_k$ and $\gJ_k^1(2)\ts
\gJ_k^e$ imply that the mapping $q\to(\vr^{(2)}\ts \gr)(q)$ is a
bijection between $\cL_k$ and $\gJ_k^1(2)\ts \gJ_k^e$, since
$U_o=\cR$ is an unitary operator  on $\cL_k$ for any $k\in \N$.

\section {  Periodic problems}
\setcounter{equation}{0}

\subsection{Periodic potentials}
We consider inverse problems on the circle. Firstly we define the
gap mapping $q\to \p=(\p_n)^\iy$ acting from $\cH$ into $ \ell^2\os
\ell^2$ from \cite{K99}. The components $\p_n\in \R^2$  are
constructed via the periodic plus Dirichlet eigenvalues plus signs
by
\[
\lb{gL1}
\begin{aligned}
&  \textstyle   \p_n=(p_{c,n},\p_{s,n})\in\R^2, \qqq
|\p_n|^2=\p_{c,n}^2+\p_{s,n}^2= {1\/4}(\l_n^+-\l_n^-)^2,
\\
&     \textstyle \p_{c,n}={1\/2}(\l_n^++\l_n^-)-\m_n,\qq
\p_{s,n}=\big||\p_n|^2- \p_{c,n}^2  \big|^{1\/2}\sign h_{s,n},\qq
h_{s,n}=\log|\vp'(1,\mu_n)|.
\end{aligned}
\]
The mapping $\p$ is a RAB between $\cH$ and $\ell^2\os \ell^2$, see
Theorem \ref{TKE} below.

 We define another gap mapping   $\gp: \cH\to \ell^2\os
\ell^2$  by $q\to \gp=(\gp_n)_1^\iy$. The components $\gp_n\in \R^2$
are constructed via the periodic plus Neumann eigenvalues plus signs
by
\[
\lb{gL2}
\begin{aligned}
&\textstyle  \gp_n=(\gp_{c,n},\gp_{s,n})\in\R^2,\qq
|\gp_n|^2=\gp_{c,n}^2+\gp_{s,n}^2= {1\/4}(\l_n^+-\l_n^-)^2,
\\
& \textstyle \gp_{c,n}={1\/2}(\l_n^++\l_n^-)-\n_n,\qqq
\gp_{s,n}=\Big||\gp_n|^2- (\gp_{c,n})^2  \Big|^{1\/2}\sign
\gh_{s,n},\qq \gh_{s,0}=\ln |\vt(1,\n_0)|.
\end{aligned}
\]

Secondly we consider inverse problems in terms of local maxima and
minima of the Lyapunov function, given by
$\D(\l)={1\/2}(\vp'(1,\l)+\vt(1,\l))$. The Lyapunov function on the
real line  has local maxima and minima at points $\l_n\in
[\l_n^-,\l_n^+]$ for all $n\in \N$, where $(-1)^n\D(\l_n^\pm)=1$ and
$(-1)^n\D(\l_n)\ge 1$. Define the corresponding mapping $h: \cH\to
\ell^2\os\ell^2$ as $h: q\to h=([2\pi n] h_n)_1^{\infty }$  from
\cite{MO75}. The components $h_n\in \R^2$ are constructed via maxima
and minima of the Lyapunov function plus Dirichlet eigenvalues plus
signs by
\[
\lb{h75}
\begin{aligned}
& h_n=(h_{c,n}, h_{s,n})\in {\R}^2,\qqq |h_n|^2
 =h_{c,n}^2+h_{s,n}^2,
 \\
& h_{c,n}=\Big||h_n|^2-h_{s,n}^2\Big|^{1\/2}{\rm sign} (\l_n-\mu_n),
 \qqq h_{s,n}=\log |\vp'(1,\mu_n)|,
 \end{aligned}
\]
here the value $|h_n|\ge 0$ is uniquely defined by the equation
$\cosh |h_n|=|\D(\l_n)|\ge 1$. Recall that $(-1)^{n}\D(\mu_n)=\cosh
h_{s,n}$ for all $n\geq 1$ and $|h_n|\ge |h_{s,n}|$, since
$(-1)^n\D$ has the maximum at $\l_n$ on the segment
$[\l_n^-,\l_n^+]$. The mapping $h$ is a RAB between $\cH$ and
$\ell^2\os \ell^2$, see below.

 We introduce similar mapping $\gh: \cH\to \ell^2\os\ell^2$ as $\gh: q\to
\gh(q)=([2\pi n]\gh_n(q))_1^{\infty }$. The components $\gh_n\in
\R^2$  are constructed via maxima and minima of the Lyapunov
function plus Neumann eigenvalues plus signs by
\[
 \lb{gh75}
\begin{aligned}
&  \gh_n=(\gh_{c,n}, \gh_{s,n})\in {\R}^2,\qq |\gh_n|=|h_n|,
  \\
& \gh_{c,n}=\Big||\gh_n|^2-\gh_{s,n}^2\Big|^{1\/2}{\rm sign}
(\l_n-\n_n),
 \qqq \gh_{s,n}=\log |\vt(1,\n_n)|.
 \end{aligned}
\]
Recall that $(-1)^{n}\D(\n_n)=\cosh \gh_{s,n}$ for all $n\geq 1$ and
$|\gh_n|\ge |\gh_{s,n}|$, since $(-1)^n\D$ has the local maximum at
$\l_n$ on the segment $[\l_n^-,\l_n^+]$. We describe well-known
results about  inverse problems on the circle.

\begin{theorem}
\lb{TKE} \no i) The mapping $h=([2\pi n]h_n)_1^\iy: \cH \to
\ell^2\os \ell^2$ given by \er{h75} is a RAB between $\cH$ and
$\ell^2\os \ell^2$. Furthermore, the following estimates hold true:
\[
\lb{esqh} \|q\| \leq 3\|h\| (6+h_+)^{1\/2}, \ \ \ \ \ \|h\| \leq
2\|q\|(1+\|q\|^{1\/3}),
\]
where $\|q\|^2=\int_0^1q^2(x)dx$ and  $\|h\|^2=\sum_{n\ge 1}|2\pi n
h_{n}|^2$ and $h_+=\sup_{n\ge 1} |h_{n}|$.

\no ii) The mapping $\p: \cH \to \ell^2\os \ell^2$ given by \er{gL1}
is a RAB between $\cH$ and $\ell^2\os \ell^2$. Furthermore, the
following estimates hold true:
\[
\lb{esqp}
 \|q\| \leq 2\|\p\|(1+\|\p\|^{1\/3}), \qqq \|\p\| \leq
 \|q\|(1+\|q\|^{1\/3}),
\]
where  $\|\p\|^2=\sum_{n\ge 1}
(\p_{c,n}^2+\p_{s,n}^2)={1\/4}\sum_{n\ge 1}|\l_n^+-\l_n^-|^2$.

\end{theorem}

\no {\bf Remark.} 1) A bijection in i) was proved in \cite{MO75}. It
was reproved in \cite{K97}, including the RAB. The proof in
\cite{K97} is simpler and is based on analytic approach from
\cite{KK97}, \cite{PT87}.

\no 2)  A bijection of $\p$ was proved in \cite{K99}. The proof is
sufficiently short, since the estimates \er{esqp} from \cite{K98}
were used. Note that there is a unique way of placing the sequence
of open tiles of lengths $\l_n^+-\l_n^-, n\ge 1$, in order on the
half line $[\l_0^+,\iy)$ so that the compliment is the set of bands
for a function $q\in \cH$, so that they are genuine gaps, see
\cite{K99}. It does not depend on the positions of the Dirichlet
spectrum $\{\m_n\}$ and $\{\sign h_{s,n}\}$.

\no 3)  The estimates \er{esqh}, \er{esqp} were obtained in
\cite{K98}, \cite{K00jde} (some estimates of $h$ were determined in
\cite{MO75}). Their proof is based on the conformal mapping theory
and trace formulas \cite{K97}.

We describe inverse problems on the circle.

 \begin{theorem}
\lb{T3}

\no i) The mappings  $\gp: \cH\to \ell^2\os \ell^2$ and $\p: \cH\to
\ell^2\os \ell^2$ given by \er{gL2}, \er{gL1} are isomorphic.
Moreover, $\gp$ is a RAB between $\cH$ and $\ell^2\os \ell^2$ and
satisfies
\[
\lb{11yy} \gp=\p\circ U_1.
\]
\no ii) The  mappings $\gh: \cH\to \ell^2\os \ell^2$  and $h: \cH
\to \ell^2\os \ell^2$ given by \er{gh75}, \er{h75} are isomorphic.
Moreover, $\gh$ is a RAB between $\cH$ and $\ell_1^2\os \ell_1^2$
 and satisfies
\[
\lb{11z} \gh=h\circ U_1.
\]
iii) Let $\s=(\s_n)_1^\iy$, where $\s_{2n}\in \{\pm 1\}$ and
$\s_{2n-1}=-1$ for all $n\in \N$.  Define replacing  mappings
$q\to\f=(\f_n)_1^\iy$ and $q\to\o=([2\pi n]\o_n)_1^\iy$, where
\[
\lb{dd1p}
\f_n=\ca \p_n \\
         \gp_n\ac,
\qqq
\o_n=\ca h_n,  \qq {\rm if} \ \s_{2n}=1\\
         \gh_n,  \qq {\rm if} \ \s_{2n}=-1 \ac.
\]
Then $\f$ is a RAB between $\cH$ and $\ell^2\os \ell^2$ and $\o$ is
a RAB between $\cH$ and $\ell^2\os \ell^2$
 and satisfy
\[
 \f=\p \circ \cU_\s, \qqq \o=h \circ \cU_\s.
\]
\end{theorem}

\no{\bf Remark.} 1) The mapping $U_1$ from this theorem is
iso-spectral for  the periodic eigenvalues, but not for the
Dirichlet and Neumann eigenvalues, see Lemma \ref{Tfx}.
\no {\bf Proof.} i) Let $q\in \cH$. Due to \er{11} 2-periodic
eigenvalues $\{\l_0^+, \l_{n}^\pm,n\ge 1\}$ are invariant under
$U_1$ and \er{12x} gives $\m_{n}(q)=\n_{n}(q^{(1)})$ for all $n\in
\N$, where $ q^{(1)}= U_1 q$. This yields
$$
\begin{aligned}
\textstyle & p_{c,n}(q)=\textstyle
\Big({\l_n^-+\l_n^+\/2}-\m_n\Big)(q)
=\Big({\l_n^-+\l_n^+\/2}-\n_n\Big)(q^{(1)})=\gp_{c,n}(q^{(1)}),
\\
\textstyle & p_{s,n}(q)=\Big(\big||p_n|^2- (p_{c,n})^2
\big|^{1\/2}\sign h_{s,n}\Big)(q)=\Big(\big||\gp_n|^2- (\gp_{c,n})^2
\big|^{1\/2}\sign \gh_{s,n}\Big)(q^{(1)})=\gp_{s,n}(q^{(1)}),
\end{aligned}
$$
since \er{32z} gives $h_{s,n}(q)=\gh_{s,n}(q^{(1)})$. Then $p=\gp
\circ U_1$ and from Theorem \ref{TKE} we deduce that $\gp$ is a RAB
between $\cH$ and $\ell^2\os \ell^2$, since due to Proposition
\ref{TPr} $U_1$ is a RAB of $\cL$ onto itself. The proof of ii) is
similar to the case i).

\no iii) Let $q^\bu= \cU_\s(q)$, where $q\in \cH$. Let $n=2j-1\ge1$
for all $j\in \N$  and $\s_n=-1$. Due to \er{U1} under the mapping
$\cU_\s$ all mixed eigenvalues satisfy
\[
\lb{qtr}    \t_n(q)=\vr_n(q^\bu), \qq \vr_n(q)=\t_n(q^\bu),
\]
and then the identities \er{E1} imply
\[
\lb{qmnq} \vp'(1,\cdot,q)=\vt(1,\cdot,q^\bu), \qq \vt(1,\cdot,q)
=\vp'(1,\cdot,q^\bu),
\]
which jointly with \er{mntz2} yields $h_{s,n}(q)=h_{s,n}(q^\bu)$ for
all $n\ge 1$.

Let $n=2j$ for all $j\in \N$. Then \er{U2}, \er{U1} yields
\[
\lb{qmn} \ca   \m_j(q)=\m_j(q^\bu),\qq \n_j(q)=\n_j(q^\bu), \qq if \
\s_{2j}=1
\\
\m_j(q)=\n_j(q^\bu),\qq \n_j(q)=\m_j(q^\bu),  \qq if \ \s_{2j}=-1\ac
.
\]
Thus  we have two cases:

\no A) If $\s_{2j}=1$, then from \er{qmn}, \er{qmnq} we have
$\f_j(q^\bu)=p_j(q^\bu)$, where
$$
h_{s,j}(q^\bu)=\ln |\vp'(1,\m_j(q^\bu),q^\bu)|=\ln
|\vp'(1,\m_j(q),q)|=h_{s,j}(q),
$$
$$
\textstyle
p_{cj}(q^\bu)=\Big({\l_n^-+\l_n^+\/2}-\m_n\Big)(q^\bu)
=\Big({\l_n^-+\l_n^+\/2}-\m_n\Big)(q)
=p_{cj}(q),
$$
$$
\textstyle p_{sj}(q^\bu)=\Big(\big||p_j|^2- (p_{c,j})^2
\big|^{1\/2}\sign h_{s,j}\Big)(q^\bu)=\Big(\big||p_j|^2- (p_{c,j})^2
\big|^{1\/2}\sign h_{s,j}\Big)(q)=p_{s,j}(q).
$$

\no B) If $\s_{2j}=-1$, then from \er{qmn}, \er{qmnq} we have
$\f_j(q^\bu)=\gp_j(q^\bu)$, where
$$
\gh_{s,j}(q^\bu)=\ln |\vt(1,\n_j(q^\bu),q^\bu)|=\ln
|\vp'(1,\m_j(q),q)|=h_{s,j}(q),
$$
$$
\textstyle \gp_{cj}(q^\bu)=\Big({\l_n^-+\l_n^+\/2}-\n_n\Big)(q^\bu)
=\Big({\l_n^-+\l_n^+\/2}-\m_n\Big)(q) =p_{cj}(q),
$$
$$
\textstyle \gp_{sj}(q^\bu)=\Big(\big||\gp_j|^2- (\gp_{c,j})^2
\big|^{1\/2}\sign \gh_{s,j}\Big)(q^\bu)=\Big(\big||p_j|^2-
(p_{c,j})^2 \big|^{1\/2}\sign h_{s,j}\Big)(q)=p_{s,j}(q).
$$
 From A and B we obtain $p=\f \circ \cU_\s$ and $\f$ is a RAB between
$\cH$ and $\ell^2\os \ell^2$, since due to Proposition \ref{TPr}
$U_\s$ is  a RAB of $\cL$ onto itself.
 The proof for mapping $\o$ is similar.
 \BBox

\subsection{Smooth periodic potentials}
We discuss isomorphic inverse problems for the case of smooth
periodic potentials from Sobolev spaces $\cH_k$ defined by
$$
\cH_k=\{q,  q^{(k)}\in \cH\},  \qq  k\ge 0.
$$

\begin{theorem}
\lb{Tpai}

i) The mappings  $\p: \cH_k\to \ell_k^2\os \ell_k^2$ and $\gp:
\cH_k\to \ell_k^2\os \ell_k^2$ given by \er{gL1}, \er{gL2}  are
isomorphic and are bijections between $\cH_k$ and $\ell_k^2\os
\ell_k^2$. They satisfy $\gp=\p\circ U_1$  and
\[
\lb{egap}
\begin{aligned}
&  \|\p\|_{k}\le C_1\|q\|_{(k)}\big(1+\|q\|_{(k)}\big)^{2s+1},
\\
& \|q\|_{(k)}\le C_2\|\p\|_{k}\big(1+\|\p\|_{k}\big)^{2m(1+s)+s},
 \end{aligned}
\]
where $\|q\|_{(k)}^2=\int_0^1|q^{(k)}|^2dx$ and
$\|\p\|_k^2=\sum_{n\ge 1} (2\pi n)^{2k}|\p_{n}|^2, \qq
|\p_{n}|={\l_n^+-\l_n^-\/2}$ and
\[
\lb{2.26x}  \tes s={2k+1\/3},\qq
m={{k+1\/3}}\big(1+tk+t^2k(k-1)+\dots +t^{k+1}k !\big), \ \ \
t={2\/3}.
\]
\no  ii) The mappings  $h: \cH_k\to \ell_{k}^2\os \ell_{k}^2$ and
$\gh: \cH_k\to \ell_{k}^2\os \ell_{k}^2$ given by \er{gL1}, \er{gL2}
are isomorphic and are bijections between $\cH_k$ and $\ell_{k}^2\os
\ell_{k}^2$. They satisfy $\gh=h\circ U_1$ and
\[
\lb{eh}
\begin{aligned}
&  \|h\|_{k}\le C_3\|q\|_{(k)}\big(1+\|q\|_{(k)}\big)^{2s+1},\qq
\\
& \|q\|_{(k)}\le C_4\|h\|_{k}(1+\|h\|_{k})^{m}(1+\sup_{j\in\N}
|h_j|)^{(k+1)(1+m)}.
 \end{aligned}
\]
The constants $C_1,.., C_4$ depend on $k$ only and
$\|h\|_k^2=\sum_{n\ge 1} (2\pi n)^{2k}|2\pi nh_{n}|^2$.

\end{theorem}

\no {\bf Remark.} The mapping $h: \cH_k\to \ell_{k}^2\os \ell_{k}^2,
k\in \N$ is a bijection, see \cite{MO75}. All estimates \er{egap},
\er{eh} are new. There is an open problem about their sharpness,
even for potentials $q\in \cH$.

\no {\bf Proof.} i) If $q\in \cH_k$, then estimates \er{egap} yield
that $\p(q)\in \ell_k^2\os \ell_k^2$. The mapping $\p: \cH_k \to
\ell_k^2\os \ell_k^2$ is an injection, since due to Theorem
\ref{TKE}, ii) the mapping $\p: \cH \to \ell^2\os \ell^2$ is a
bijection.

Let $b\in \ell_k^2\os \ell_k^2$. Then due to Theorem \ref{TKE}, ii)
there exists a unique $q\in \cH_0$, such that $\p(q)=b$. Thus using
the following results:  if $q\in \cH$ and $k\in \N$, then
\[
\lb{2L} (|\p_n|)_1^\iy\in \ell_k^2 \qq  {\rm or} \qq h\in \ell_{k}^2
\qqq \Leftrightarrow  \qqq q\in \cH_k.
\]
see e.g., Corollary 3.4 in \cite{MO75} or \cite{K06}, we deduce that
$q\in \cH_k$.

We show  \er{egap}. We need estimates  (see Theorem 2.3 and Theorem
2.6 from \cite{K06})
\[
\lb{2.9} Q_k\le c_1^2\|\p\|_{k}^2\big(1+\|\p\|_{k}^{2s}\big), \tes
\]
\[
\lb{2.10} \|\g\|_{k}^2\le c_2^2Q_k\big(1+ Q_k\big),
\]
where $Q_{k}\ge 0$ is some non-linear functional of $q, q',...,
q^{(k)}$ and
\[
\lb{2.24} Q_k\le c_3^2\|q^{(k)}\|^{2}\big(1+\|q^{(k)}\|\big)^{2s},
\]
\[
\lb{2.25} \|q^{(k)}\|\le c_4^2 Q_k^{1\/2}\big(1+Q_k^{m}\big),
\]
for some constants $c_1,..,c_4$ depending on $k$ only. Let
$A=Q_k^{1\/2}\ge 0$. Due to an inequality $(1+x)^{1\/2}\le
1+x^{1\/2}$ for $x\ge 0$ we rewrite these estimates in the  form:
 \[
 \lb{2.9e}
 A\le c_1\|\p\|_{k}\big(1+\|\p\|_{k}\big)^s,
\]
\[
\lb{2.10e} \|\p\|_{k}\le c_2A\big(1+ A\big),
\]
and
\[
\lb{2.24e} A\le c_3\|q^{(k)}\|\big(1+\|q^{(k)}\|\big)^s,
\]
\[
\lb{2.25e} \|q^{(k)}\|\le c_4 A\big(1+A^{2m}\big),
\]
\no $\bu$ Let $A<1$. Then \er{2.10e}, \er{2.25e} and \er{2.9e} yield
\[
\lb{h1}
\begin{aligned}
  \|\p\|_{k}\le 2c_2 A\le
  2c_2c_3\|q^{(k)}\|\big(1+\|q^{(k)}\|\big)^s,
\\
 \|q^{(k)}\|\le 2c_4 A\le 2c_4
 c_1\|\p\|_{k}\big(1+\|\p\|_{k}\big)^s.
 \end{aligned}
\]
$\bu$ Let $A\ge 1$. Then \er{2.10e}, \er{2.25e} and \er{2.9e}  yield
\[
\lb{h2}
\begin{aligned}
  \|\p\|_{k}\le 2c_2A^2 \le
  2c_2c_3^2\|q^{(k)}\|^2\big(1+\|q^{(k)}\|\big)^{2s}\le
  2c_2c_3^2\|q^{(k)}\|\big(1+\|q^{(k)}\|\big)^{2s+1},
\\
 \|q^{(k)}\|\le 2c_4 A^{1+2m}\le 2c_4
 c_1^{1+2m}\|\p\|_{k}\big(1+\|\p\|_{k}\big)^{s(1+2m)+2m}.
 \end{aligned}
\]
Estimates \er{h1}, \er{h2} imply \er{egap}.

\no ii) The proof for mappings $h$ and $\gh$ is similar, we need
only to show \er{eh}. Let $q\in \cH_{k}, k\in \N$. We need the
following estimates (see Theorem 2.1 from \cite{K06}):
\[
\lb{2.1h}
\begin{aligned}
& \|h\|_{k}^2\le b^2Q_k (1+Q_k),\ \ \ b^2=4^{8k+11},
\\
& Q_k\le {1\/\pi (k+1)}h_+^{2(k+1)} \|h\|_{k}^2,\qqq
h_+=\sup_{j\in\N} |h_j|.
\end{aligned}
\]
Due to an inequality $(1+A^2)^{1\/2}\le 1+A$ for $A=Q_k^{1\/2}\ge 0$
we rewrite \er{2.1h} in the  form:
\[
\lb{h3}
\begin{aligned}
& \|h\|_{k}\le b A (1+A),
\\
& A\le \s_* \|h\|_{k}h_+^{k+1},  \qqq \s_*=(\pi (k+1))^{-{1\/2}}.
\end{aligned}
\]
\no $\bu$ Let $A<1$. Then \er{h3}, \er{2.24e}  and  \er{2.25e} yield
\[
\lb{h4}
\begin{aligned}
  \|h\|_{k}\le 2b A\le
  2bc_3\|q^{(k)}\|\big(1+\|q^{(k)}\|\big)^s,
\\
 \|q^{(k)}\|\le 2c_4 A\le 2c_4 \s_*\|h\|_{k}h_+^{k+1}.
 \end{aligned}
\]
$\bu$ Let $A\ge 1$. Then \er{h3}, \er{2.24e}  and  \er{2.25e}  yield
\[
\lb{h5}
\begin{aligned}
  \|h\|_{k}\le 2b A^2\le
  2bc_3^2\|q^{(k)}\|^2\big(1+\|q^{(k)}\|\big)^{2s}\le
  2bc_3^2\|q^{(k)}\|\big(1+\|q^{(k)}\|\big)^{1+2s},
\\
 \|q^{(k)}\|\le 2c_4 A^{1+m}\le 2c_4
 \s_*^{1+m}\|h\|_{k}^{1+m}h_+^{(k+1)(1+m)}.
 \end{aligned}
\]
Estimates \er{h4}, \er{h5} imply \er{eh}. \BBox


\section {Appendix: asymptotics and  trace formulas}
\setcounter{equation}{0}

\subsection{Asymptotics}
We discuss asymptotics of fundamental solutions.

\begin{proposition} \lb{Tck09}
Let $q\in \cL$. Then following asymptotics as $n\to \iy$ hold true:
\[
\lb{ck1} \ln \big[ (-1)^n2\m_n^o\dot\vp(1,\m_n) \big]=\ell_{1}^2(n),
\]
\[
\lb{ck2} \ln \big[ (-1)^n2\sqrt{\t_n^o}\ \dot\vp'(1,\t_n)
\big]=\ell_{1}^2(n),
\]
\[
\lb{ck3} \ln \big[ (-1)^n2\sqrt{\vr_n^o}\ \dot\vt(1,\vr_n)
\big]=\ell_{1}^2(n).
\]

\end{proposition}

\no{\bf Proof.} Asymptotics \er{ck1} was proved in \cite{CK09}. We
show \er{ck2}. Recall that the sequence $\hat{\t}_n=\t_n-\t_n^o,
n\ge 1$ belongs to $\ell^2$. Using \er{E1} and
$\dot\vp_o'(1,\t_n^o)={(-1)^n\/2\sqrt{\t_n^o}}$ at $q=0$ we obtain
\[
\dot \vp'(1,\t_n)={-1\/\t_n^o}\prod_{j\ne
n}^{+\iy}\frac{\t_j\!-\!\t_n}{\t_j^o}
= {(-1)^n\!\/2\sqrt{\t_n^o} }\prod_{j\ne
n}^{+\iy}\frac{\t_j\!-\!\t_n}{\t_j^o-\t_n^o}
= {(-1)^n\/2\sqrt{\t_n^o} }\prod_{j\ne n}^{+\iy}\lt[1+\frac{\hat
\t_j-\hat\t_n}{\t_j^o-\t_n^o}\rt].
\]
Then
\[
\lb{fS}
\begin{aligned}
& \log(-1)^n2\sqrt{\t_n^o} \dot \vp'(1,\t_n)=\log\prod_{j\ne
n}^{+\iy}\lt[1+\frac{\hat \t_j-\hat\t_n}{\t_j^o-\t_n^o}\rt] =
\sum_{j\ne n}^{+\iy}\lt[\frac{\hat \t_j-\hat\t_n}{\t_j^o-\t_n^o}+
  \frac{O(1)}{(\t_j^o-\t_n^o)^2} \rt]
\\
& =\sum_{j\ne n}^{+\iy}\frac{\hat
\t_j-\hat\t_n}{\t_j^o-\t_n^o}+\frac{O(1)}{n^2}=\sum_{j\ne
n}^{+\iy}\frac{\hat
\t_j}{\t_j^o-\t_n^o}-{\hat\t_n\/4\t_n^o}+\frac{O(1)}{n^2}=\sum_{j\ne
n}^{+\iy}\frac{\hat \t_j}{\t_j^o-\t_n^o}+\frac{O(1)}{n^2},
\end{aligned}
\]
since due to $\frac{1}{\t_j^o-\t_n^o}={1\/2\pi
\sqrt{\t_n^o}}\big({1\/j-n}-{1\/j+n-1}\big)$ and
${1\/2n-1}={\pi\/2\sqrt{\t_n^o}}$   we have
\[
\lb{qw1}
\begin{aligned}
 \sum_{j\ne
n}^{+\iy}\frac{1}{\t_j^o-\t_n^o}=\lim_{m\to \iy} \sum_{j\ne
n}^{m}\frac{1}{\t_j^o-\t_n^o}={1\/2\pi \sqrt{\t_n^o}}\ \lim_{m\to
\iy}A_m, \qq A_m=\sum_{j=1,j\ne
n}^{m}\textstyle\big({1\/j-n}-{1\/j+n-1}\big),
\\
A_m={1\/2n-1}+\sum_{j=-m,j\ne n}^{m} {1\/j-n}= {\pi\/2
\sqrt{\t_n^o}}+o(1)\qq \as \qq m\to \iy.
\end{aligned}
\]
Let $\hat \t=(\hat \t_j)_{j\in \Z}$, where $\hat \t_{\pm j}={\hat
\t_j}$ for $j\in \N$ and $\hat \t_0=0$. The last sum in \er{fS}
satisfies
\[
\lb{fSqq}
\begin{aligned}
\sum_{j\ne n}^{+\iy}\frac{\hat \t_j}{\t_j^o-\t_n^o}={1\/2\pi
\sqrt{\t_n^o}}\sum_{j\ne n}^{+\iy}\big({\hat \t_j\/j-n}-{\hat
\t_j\/j+n-1}\big)
\\
={1\/2\pi \sqrt{\t_n^o}}\rt[ {\pi \hat
\t_n\/2\sqrt{\t_n^o}}+\sum_{j\in \Z\sm\{n\}}{\hat \t_j\/j-n} \rt]=
{\hat \t_n\/4\t_n^o}+{1\/2\pi \sqrt{\t_n^o}}(\cF \hat \t)_n \qqq
\end{aligned}
\]
where $\cF$ is the linear operator on $\ell^2(\Z)$ and given by
$(\cF \hat \t)_n=\sum_{j\in \Z\sm\{n\}}{\hat \t_j\/j-n}$. Due to the
identity $\sum_{j\ne 0}{e^{i2\pi jt}\/j}=-i(t-{1\/2}), t\in (0,1)$
and the Fourier transform we deduce that $\cF$ is the bounded
operator in $\ell^2(\Z)$. Thus jointly  \er{fS}, \er{fSqq} it gives
\er{ck2}.

We show \er{ck3}. Let $q^o=U_o(q)$.   From   \er{32o} we obtain
$\vt(1,\cdot,q)=\vp'(1,\cdot,q^o)$. Then \er{m4o} gives
$\vt(1,\vr_n(q),q)=\vp'(1,\t_n(q^o),q^o)$ and Theorem \ref{TMO} iii)
implies \er{ck3}. \BBox

In order to determined asymptotics of norming constants we need
asymptotics of fundamental solutions.  Recall that a solution of
$-y''+qy=z^2 y, z>1$ has the form
\[
\lb{Y1}
\begin{aligned}
\tes y(x,z)=e^{izx}Y(x,z),\qq
Y(x,z)=1+\sum_1^{k}{u_j(x)\/\vs^j}+{\wt u_k(x,z)\/\vs^{k+1}}, \qq
\vs=i2z,
 \end{aligned}
\]
see Sect. 3 in \cite{MO75},  where $u_1(x)=\int_0^x q(t)dt,....$ and
\[
\lb{Y2}
\begin{aligned}
\wt u_k\in \cL_2,\qq \wt u_k(0,z)=\wt u_k'(0,z)=u_j(0)=0, \qq
u_{j}\in \cL_{2},\qq j\in
 \N_{k+1},
 \end{aligned}
\]
\[
\lb{Y3}
\begin{aligned}
 \wt u_k(1,z)=u_{k+1}(1)-q_*^{k}(z)+{\tes{1\/\vs}}(C_q+c_qe^{-i2z}+\wh
K(z)), \qq  c_q, C_q\in \R,
\\
q_*^k(z)=(-1)^ke^{-i2z}\int_0^1 e^{i2zt}q^{(k)}(t)dt,\qq \wh
K(z)=\int_0^1 e^{-i2zt}g(t)dt,\qq K\in L^2(0,1),
 \end{aligned}
\]
We can rewrite $\vp, \vt$ in  terms of $y(x,\pm z)$. For example, we
compute the identities
\[
\lb{Y4}
\begin{aligned}
& \tes \vt(1,z)={1\/w(z)}(y_0'(z)y_1(-z)-  y_0'(-z)y_1(z)),\qqq
\vp(1,z)={1\/w(z)}(y_1(z)-y_1(-z)),
\\
& {\where} \qq y_0'(z)=y'(0,z),\qq y_1(z)=y(1,z), \qq
w(z)=y_0'(z)-y_0'(-z).
 \end{aligned}
\]
We rewrite the function $y_0'(z)$, the Wronskian $w(z)$ due to
\er{Y1}, \er{Y2} in terms of $1/\vs$:
\[
\lb{y4}
\begin{aligned}
& \tes y_0'(z)=iz+\sum_1^k{u_j'(0)\/\vs^j}={\vs}y_\cd(z),\qq
y_\cd(\vs)={1 \/2}+\sum_1^k{2u_j'(0)\/\vs^{j+1}},
\\
& \tes w(z)=\vs+\sum_1^k{u_j'(0)\/\vs^j}\big(1-(-1)^j\big)=\vs
w_\cd(\vs),\qq w_\cd(\vs)=1+\sum_{2\le 2j\le k+1}  {\tes
{2u_{2j-1}'(0)\/\vs^{2j}}}.
 \end{aligned}
\]

 \begin{lemma}
\lb{Tafs} Let  $q\in \cL_k, k\in \N$. Then the norming constants
$h_{s,n}=\ln |\vp'(1,\m_n,q)|$ satisfy
\[
\lb{y}
\begin{aligned}
h_{s,n}=\sum_{1\le j\le d}\wt\vp_j \ve^{2j} +E_n
\ve^{k+1}+\ell_{k+2}^2(n)\qq \as \qq n\to \iy,
 \end{aligned}
\]
where $\ve={1\/2\pi n}$,  the coefficients  $\wt\vp_j\in \R,\  j\in
\N_d, \ d=\big[{k+1\/2}\big]$ depend on $q$ and
\[
\lb{yc}
\begin{aligned}
E_n=(-1)^{1+[{k\/2}]} \int_0^1 q^{(k)}(x)\cG_k(nx)dx, \qq
\cG_k(nx)=\ca \sin[2\pi nx], \qq k\in 2\N
\\  \cos[2\pi nx], \qq k+1\in 2\N \ac .
 \end{aligned}
\]

 \end{lemma}

\no {\bf Proof.} From \er{Dm1} we deduce that the eigenvalues
$z_n=\sqrt{ \m_n}$ have asymptotics
\[
\lb{y5}
\begin{aligned}
& z_n=\pi n+r_n,\qqq r_n=\sum_{1\le j\le d}a_j\ve ^{2j+1} +\ve
^{k+1}\wt r_n, \qqq  (\wt r_n)_1^\iy\in \ell^2,
\\
& z_n=\pi n{\bv_n}, \qqq  \bv_n=1+2r_n\ve=1+\sum_{1\le j\le
d}2a_j\ve ^{2j+2} +\ell_{k+2}^2(n), \qqq
\end{aligned}
\]
where $(a_j)_1^d\in \R^{d}$. Moreover, the function
$\bv_n=1+2r_n\ve$ and $1/(i2z_n)$ satisfy
\[
\lb{y6}
\begin{aligned}
{1\/\bv_n^s}={1\/(1+2r_n\ve)^s}=1+\sum_{1\le j\le d}c_j(s)\ve
^{2j+2} +\ell_{k+2}^2(n),
\\
{1\/(i2z_n)^{s}}={(-i)^s \ve^{s}\/\bv_n^{s}}=(-i)^s\sum_{1\le j\le
d}C_j(s)\ve ^{2j+2+s} +\ell_{k+1+s}^2(n), \qq \forall \ s\in \N,
 \end{aligned}
\]
where $C_j(s), c_j(s)$ are polynomial of $a_j$. The asymptotics of
$(i2z_n)^{-s}$ has two terms. The first is $\ve^s \P$, where $\P$ is
a polynomial in $\ve$ with even power $2j+2\le d+2$. The second term
is a remainder $\ell_{k+1+s}^2(n)$. It is a crucial fact in our
proof. Using \er{y4}, \er{y5},\er{y6}  we obtain
\[
\lb{y7}
\begin{aligned}
& w_\cd(z_n)=1+ \sum_{1\le j\le d} \wt w_j\ve ^{2j}
+\ell_{k+3}^2(n),\qqq y_\cd(z_n)={\tes{1\/2}}+ \sum_{1\le j\le 2k}
\wt y_{j}\ve ^{j} +\ell_{k+3}^2(n),
\\
& {1\/w_\cd(z_n)}=1+ \sum_{1\le j\le d}\tes \wh w_j\ve ^{2j}
+\ell_{k+3}^2(n),
 \end{aligned}
\]
for some coefficients $\wt w_j, \wt y_{j}, \wh w_j$  and thus
\er{Y1}-\er{Y3} imply
\[
\lb{y8}
\begin{aligned}
 Y(1,z_n)=1+\sum_1^k{u_j(1)\/(2iz_n)^j}+{\wt
u_k(1,z)\/(2iz_n)^{k+1}}=1+\sum_1^{k+2} \gp_j\ve^j+\ve^{k+1}
q_{k}^\bu(n)+\ell_{k+2}^2(n),
\end{aligned}
\]
for some coefficients $\gp_j$ depending on $q$.

From the Wronskian we get $\vt(1,\m_n)\vp'(1,\m_n)=1$. Then in order
to get \er{y}  we can study the asymptotics of
$e^{h_{s,n}}=(-1)^n\vt(1,\m_n)$ given by \er{Y4}. From \er{Y4},
\er{Y1}, \er{y8} we have
\[
\lb{y9}
\begin{aligned}
e^{h_{s,n}}={y_0'(z)e^{-iz}Y_{1}(-z)-
y_0'(-z)e^{iz}Y_{1}(z)\/(-1)^nw(z)}\Big|_{z_n}
={y_\cd(z)e^{-ir_n}Y_{1}(-z)+
y_\cd(-z)e^{ir_n}Y_{1}(z)\/w_\cd(z)}\Big|_{z_n}
\\
=\sum_{m=0}^{2k}{i^m\/m!}
A_m(z_n)+O(r_n^{2k+1})=\sum_{m=0}^{2k}{i^m\/m!}
A_m(z_n)+O(\ve^{6k+3}),
 \end{aligned}
\]
since due to \er{y5} we have   $r_n=O(\ve^{3})$ and where
\[ \lb{y10}
\begin{aligned}
A_m(z_n)={r_n^m\/w_\cd(z_n)}B_m(z_n), \qqq B_m(z)=(-1)^m
y_\cd(z)Y_{1}(-z)+ y_\cd(-z)Y_{1}(z).
 \end{aligned}
\]
Consider the main term $A_0(z)$. Since $B_0(z)$ is even in $z$, then
\er{y7}, \er{y8} give
\[
\lb{y11}
\begin{aligned}
& B_0(z_n)=y_\cd(z_n)Y_{1}(-z_n)+y_\cd(-z_n)Y_{1}(z_n)=1+\sum_{1\le
j\le d}b_j\ve ^{2j} +E_n\ve^{k+1}+\ell_{k+2}^2(n),
\\
& A_0(z_n)={B_0(z_n)\/w_\cd(z_n)}=1+\sum_{1\le j\le d}\wt b_j\ve
^{2j} +E_n \ve^{k+1}+\ell_{k+2}^2(n),
  \end{aligned}
\]
where $b_j, \wt b_j$ polynomials from  $a_j, u_j'(0), u_j(1), j\in
\N_k $ and $E_n=-{q_{k}^\bu(n)+(-1)^{k+1}q_{k}^\bu(-n)\/i^{k+1}2}$
satisfies
\[
\lb{y11zx}
\begin{aligned}
{\rm if}\ {\tes {k\/2}}\in \N \ \Rightarrow E_n= {\int_0^1
q^{(k)}(t)(e^{i2\pi nt}- e^{-i2\pi nt} )dt\/(-1)^{1+{k\/2}} 2i} =
(-1)^{1+{k\/2}}\int_0^1 q^{(k)}(t)\sin [2\pi nt]dt,
  \end{aligned}
\]
which yields \er{yc}  for even $k$. Similar arguments imply \er{yc}
for odd $k$.

Consider $A_m(z)$ with even $m\ge 2$. Then  the asymptotics
\er{y11}, \er{y5} imply
\[
\lb{y12}
\begin{aligned}
A_m(z_n)=r^m A_0(z)=r^m \Big(1+\sum_{1\le j\le d}\wt b_j\ve ^{2j}
+\ell_{k+1}^2(n)\Big) =\sum_{3\le j\le 1+d+3m}b_j'\ve ^{2j}
+\ell_{k+1+3m}^2(n).
  \end{aligned}
\]
Thus $A_m(z_n)$ is an even polynomial of $\ve$ with order $\le
2(1+d+3m)$ plus a reminder $\ell_{k+1+3m}^2(n)$.

\no Let $m$ be odd. Then
$B_m(z)=y_\cd(-z)Y_{1}(z)-y_\cd(z)Y_{1}(-z)$ is odd in $z$. Then
\er{y8}, \er{y7} give
$$
\begin{aligned}
& B_1(z_n)=y_\cd(-z)Y_{1}(z_n)-y_\cd(z_n)Y_{1}(-z_n)=\sum_{1\le
2j+1\le d}t_{j}\ve ^{2j+1} +\ell_{k+1}^2(n),
\\
& {B_1(z_n)\/w_\cd(z_n)}={1\/w_\cd(z_n)} \Big(\!\!\sum_{1\le 2j+1\le
d}\!\!\!\! t_{j}\ve ^{2j+1} +\ell_{k+1}^2(n)\Big)=\!\!\sum_{1\le
2j+1\le d}\wt t_{j}\ve ^{2j+1} +\ell_{k+1}^2(n),
  \end{aligned}
$$
where the constants  $t_j, \wt t_j$ are polynomials from  $a_j,
u_j'(0), u_j(1), j\in \N_k $, and using \er{y5} we obtain
\[
\lb{y11zzz}
\begin{aligned}
A_m(z_n)=r_n^m{B_1(z)\/w_\cd(z)} =r_n^m \Big(\!\!\sum_{1\le j\le
d}\!\!\wt t_j\ve ^{2j+1} +\ell_{k+1}^2(n)\Big) =\!\!\sum_{3\le j\le
d+3m}\!\!t_j'\ve ^{2j} +\ell_{k+2}^2(n),
  \end{aligned}
\]
for some constants  $t_j'$. Collecting estimates \er{y11}, \er{y12},
\er{y11zzz} we obtain
$$
e^{h_{s,n}}=1+P(\ve)+\ve^{k+1}E_n+\ell_{k+2}^2(n),
$$
where $P(\ve)$ is a polynomial in $\ve$ with even power $2j+2\le
d+2$, which yields  \er{y}.\BBox

 \begin{lemma}
\lb{Tm1} Let  $q\in \cL_k, k\in \N$. Then the norming constants
$\gt_n=\ln |\vp(1,\t_n)\sqrt{\t_n^o}|$ satisfy
\[
\lb{my}
\begin{aligned}
\gt_n=\sum_{1\le j\le d}\gt_{j,n} \d^{2j} +E_n
\ve^{k+1}+\ell_{k+2}^2(n)\qq \as \qq n\to \iy,
 \end{aligned}
\]
where $\d={1\/2\pi (n-{1\/2})}$,  the coefficients  $\gt_{j,n}\in
\R,\ j\in \N_d, \ d=\big[{k+1\/2}\big]$ depend on $q$ and
\[
\lb{myc}
\begin{aligned}
E_n=(-1)^{1+[{k\/2}]} \int_0^1 q^{(k)}(x)\cG_k(n'x)dx, \qq
\cG_k(nx)=\ca \sin[2\pi nx], \qq k\in 2\N
\\  \cos[2\pi nx], \qq k+1\in 2\N \ac .
 \end{aligned}
\]
 \end{lemma}

\no {\bf Proof.} From \er{Dm2} we deduce that the eigenvalues
$z_n=\sqrt{ \t_n}$ have asymptotics
\[
\lb{my5}
\begin{aligned}
& z_n=\pi n'+r_n,\qq {\tes n'=n-{1\/2} }, \qq r_n=\sum_{1\le j\le
d}a_j\d^{2j+1} +\d ^{k+1}\wt r_n, \qqq  (\wt r_n)_1^\iy\in \ell^2,
\\
& z_n=\pi n'{\bv_n}, \qqq  \bv_n=1+2r_n\d=1+\sum_{1\le j\le d}2a_j\d
^{2j+2} +\ell_{k+2}^2(n), \qqq
\end{aligned}
\]
where $(a_j)_1^d\in \R^{d}$. Moreover, the function $\bv_n=1+2r_n\d$
and $\vs=i2z$ satisfies
\[
\lb{my6}
\begin{aligned}
{1\/\bv_n^s}={1\/(1+2r_n\d)^s}=1+\sum_{1\le j\le d}c_j(s)\d ^{2j+2}
+\ell_{k+2}^2(n),
\\
{1\/(i2z_n)^{s}}={(-i)^s \d^{s}\/\bv_n^{s}}=(-i)^s\sum_{1\le j\le
d}C_j(s)\d^{2j+2+s} +\ell_{k+1+s}^2(n), \qq \forall \ s\in \N,
 \end{aligned}
\]
where $C_j(s), c_j(s)$ are polynomial of $a_j$. The asymptotics of
$z_n^{-s}$ has two terms. The first is $\ve^s \cP$, where $\cP$ a
polynomial in $\ve$ with even power $2j+2\le d+2$. The second term
is a remainder $\ell_{k+1+s}^2(n)$. It is a crucial fact in our
proof.

Recall  that  due to \er{Y1} a solution to $-y''+qy=z^2 y, z>1$ has
the form
\[
\lb{my1}
\begin{aligned}
\tes y(x,z)=e^{izx}Y(x,z),\qq Y(x,z)=1+\sum_1^k{u_j(x)\/\vs^j}+{\wt
u_k(x,z)\/\vs^{k+1}}, \qq \vs=i2z.
 \end{aligned}
\]
We need also the Wronskian $w(z)=\{y(\cdot,-z),y(\cdot,z)
\}=y_0'(z)-y_0'(-z)$, where the function  $y_0'(z)=y'(0,z)$.
  The function $\vp$ has the form
\[
\lb{my3}
\begin{aligned}
\tes \vp(1,z)={1\/w(z)}(y_1(z)-y_1(-z)),\qqq \where \ y_1(z)=y(1,z).
 \end{aligned}
\]
 Using \er{y4}, \er{y5},\er{y6}  we obtain
\[
\lb{my7}
\begin{aligned}
& w_\cd(z)=1+ \sum_{1\le j\le d} \wt w_j\d ^{2j}
+\ell_{k+3}^2(n),\qqq y_\cd(z)={\tes{1\/2}}+ \sum_{1\le j\le 2k} \wt
y_{j}\d^{j} +\ell_{k+3}^2(n),
\\
& {1\/w_\cd(z)}=1+ \sum_{1\le j\le d}\tes \wh w_j\d^{2j}
+\ell_{k+3}^2(n),
 \end{aligned}
\]
for some coefficients $\wt w_j, \wt y_{j}, \wh w_j$  and
\er{Y1}-\er{Y3} imply
\[
\lb{my8x}
\begin{aligned}
 Y(1,z_n)=1+\sum_1^k{u_j(1)\/(2iz_n)^j}+{\wt
u_k(1,z)\/(2iz_n)^{k+1}}=1+\sum_1^{k+1} \gp_j\d^j+\d^{k+1}
q_{k}^\bu(n)+\ell_{k+2}^2(n),
\\
q_*^k(z)=(-1)^{k+1}\int_0^1 e^{i2\pi n'x}q^{(k)}(x)dx
\end{aligned}
\]
for some coefficients $\gp_j$ depending on $q$. From \er{y5},
\er{my3}, \er{my8x} we have that the norming constant
$e^{\gt_n}=(-1)^n (\pi n')\vp(1,z_n)$ satisfies
\[
\lb{my9}
\begin{aligned}
e^{\gt_n}=(\pi n'){e^{-iz_n}Y_{1}(-z_n)-e^{iz_n}Y_{1}(z_n)\/(-1)^n
(2iz_n) w_\cd(z_n)} ={e^{-ir_n}Y_{1}(-z_n)+ e^{ir_n}Y_{1}(z_n)\/2
\bv_n w_\cd(z_n)}
\\
=\sum_{m=0}^{2k}{i^m\/m!}
A_m(z_n)+O(r_n^{2k+1})=\sum_{m=0}^{2k}{i^m\/m!}
A_m(z_n)+O(\d^{6k+3}),
 \end{aligned}
\]
since due to \er{my5} we have   $r_n=O(\d^{3})$ and where
\[ \lb{my10}
\begin{aligned}
A_m(z_n)={r^m\/\bv_n w_\cd(z_n)}B_m(z_n), \qqq
B_m(z_n)={(-1)^mY_{1}(-z_n)+Y_{1}(z_n)\/2}.
 \end{aligned}
\]
Consider the main term $A_0(z_n)$. Since $B_0(z)$ is even in $z$,
then \er{y7}, \er{my8x} give
\[
\lb{my11}
\begin{aligned}
& B_0(z_n)={Y_{1}(-z_n)+Y_{1}(z_n)\/2}=1+\sum_{1\le j\le d}b_j\d
^{2j} +E_n\d^{k+1}+\ell_{k+2}^2(n),
\\
& A_0(z_n)={B_0(z)\/\bv_n w_\cd(z)}=1+\sum_{1\le j\le d}\wt b_j\d
^{2j} +E_n \d^{k+1}+\ell_{k+2}^2(n),
  \end{aligned}
\]
where $b_j, \wt b_j$ polynomials from  $a_j, u_j'(0), u_j(1), j\in
\N_k $ and $E_n=-{q_{k}^\bu(n)+(-1)^{k+1}q_{k}^\bu(-n)\/i^{k+1}2}$
satisfies
\[
\lb{my11zx}
\begin{aligned}
{\rm if}\ {\tes {k\/2}}\in \N \ \Rightarrow E_n= {\int_0^1 (e^{i2\pi
nt}- e^{-i2\pi nt} )q^{(k)}(t)dt\/(-1)^{1+{k\/2}} i2}=
(-1)^{1+{k\/2}}\int_0^1 \sin [2\pi nt]q^{(k)}(t)dt,
  \end{aligned}
\]
which yields \er{yc}  for even $k$. Similar arguments imply \er{yc}
for odd $k$.

Consider $A_m(z)$ with even $m\ge 2$. Then  the asymptotics
\er{y11}, \er{y5} imply
\[
\lb{my12}
\begin{aligned}
A_m(z_n)=r^m A_0(z)=r^m \Big(1+\sum_{1\le j\le d}\wt b_j\d^{2j}
+\ell_{k+1}^2(n)\Big) =\sum_{3\le j\le 1+d+3m}b_j'\d^{2j}
+\ell_{k+1+3m}^2(n).
  \end{aligned}
\]
Thus $A_m(z_n)$ is an even polynomial of $\ve$ with order $\le
2(1+d+3m)$ plus a reminder $\ell_{k+1+3m}^2(n)$.

Let $m\in \N$ be odd. Then
$B_1(z)=y_\cd(-z)Y_{1}(z)-y_\cd(z)Y_{1}(-z)$ is odd in $z$ and
\er{my6}, \er{my8x}, \er{my7} give
$$
\begin{aligned}
& B_1(z_n)=y_\cd(-z)Y_{1}(z_n)-y_\cd(z_n)Y_{1}(-z_n)=\sum_{1\le
2j+1\le d}t_{j}\d^{2j+1} +\ell_{k+1}^2(n),
\\
& {B_1(z_n)\/\bv_n  w_\cd(z_n)}={1\/\bv_n  w_\cd(z_n)}
\Big(\!\!\!\!\sum_{1\le 2j+1\le d}\!\!\!\! t_{j}\d^{2j+1}
+\ell_{k+1}^2(n)\Big)=\!\!\!\!\sum_{1\le 2j+1\le d}\wt
t_{j}\d^{2j+1} +\ell_{k+1}^2(n),
  \end{aligned}
$$
where the constants  $t_j, \wt t_j$ are polynomials from  $a_j,
u_j'(0), u_j(1), j\in \N_k $, and using \er{my5} we obtain
\[
\lb{my11zzz}
\begin{aligned}
A_m(z_n)=r_n^m{B_1(z_n)\/\bv_n  w_\cd(z)} = r_n^m
\Big(\!\!\!\sum_{1\le j\le d}\!\!\wt t_j\d^{2j+1}
+\ell_{k+1}^2(n)\Big) =\!\!\sum_{3\le j\le d+3m}\!\!t_j'\d^{2j}
+\ell_{k+2}^2(n),
  \end{aligned}
\]
for some constants  $t_j'$. Collecting estimates \er{my11},
\er{my12}, \er{my11zzz} we obtain
$$
e^{h_{s,n}}=1+P(\ve)+\ve^{k+1}E_n+\ell_{k+2}^2(n),
$$
where $P(\ve)$ is a polynomial in $\ve^2$ with even power $2j+2\le
d+2$, which yields \er{my}.\BBox


\subsection{\bf  Trace formulas }
We discuss trace formulas for Sturm-Liouvill problem
 for $q\in \cL_1$.

\no $\bu $ Gel'fand and Levitan determined trace formulas
\cite{GL53} for Dirichlet and Neumann eigenvalues
\[
\lb{tr1}
\begin{aligned}
 \textstyle    {q(0)+q(1)\/4}=\sum_{n\ge 1} (\m_n^o-\m_n),
 \\
 \textstyle    {q(0)+q(1)\/4}=\sum_{n\ge 0} (\n_n-\n_n^o),
 \end{aligned}
\]
and mixed  eigenvalues
\[
\lb{tm1}
 \textstyle    {q(0)-q(1)\/4}=\sum_{n\ge 1} (\t_n^o-\t_n),
 \]
\[
\lb{tm2}
  \textstyle    {q(0)-q(1)\/4}=\sum_{n\ge 1} (\vr_n-\t_n^o).
\]
$\bu $  Magnus and Winkler \cite{MW66} determined trace formulas for
periodic case:
\[
\lb{tr3}
\begin{aligned}
 \textstyle
 0=\sum\limits_{n\in\N} (\l_{2n-1}^++\l_{2n-1}^--2\m_{2n-1}^o),
 \\
 \textstyle   \l_{0}^+=-\sum_{n\ge 1}
 (\l_{2n}^++\l_{2n}^--2\m_{2n}^o).
 \end{aligned}
\]
$\bu $ Another trace formula for periodic case was obtained  (see
e.g., \cite{IM75}, \cite{T77})
\[
\lb{tr4}
\begin{aligned}
 \textstyle
   q(0)=\l_{0}^++\sum_{n\ge1} (\l_{n}^++\l_n^--2\m_n),\qq q\in \cH_1.
 \end{aligned}
\]
In fact summing \er{tr1} for Dirichlet eigenvalues and \er{tr3} we
obtain \er{tr4}. Similar arguments imply the trace formula for
periodic case via the Neumann eigenvalues in \er{tr1}:
\[
\lb{tr5}
\begin{aligned}
 \textstyle
   q(0)=2\n_0-\l_{0}^+-\sum_{n\ge1} (\l_{n}^++\l_n^--2\n_n).
 \end{aligned}
\]

We recall the known  facts from the theory of Fourier series.

 \begin{lemma}
\lb{Tfc} Let $q=q_{od}+q_{ev}$ for some  $q\in \cL_1$, where
\[
\lb{q1}
\begin{aligned}
 \textstyle
 q_{od}(x)=\sum_{n\ge 1} 2q_{c,2n-1}\cos \pi(2n-1)x,\qq
 q_{ev}(x)=\sum_{n\ge 1} 2q_{c,2n}\cos \pi 2nx,\qq x\in (0,1),
 \end{aligned}
\]
and $q_{c,n}=\int_0^1q(x)\cos \pi nxdx$. Then the following formulas
hold true:
\[
\lb{f3}
\begin{aligned}
 \textstyle
q_{ev}(0)={q(0)+q(1)\/2}, \qqq    q_{od}(0)={q(0)-q(1)\/2}.
 \end{aligned}
\]
 \end{lemma}

\no {\bf Proof.} There is a Fourier series $q(x)=\sum_{n\ge 1}
2q_{c,n}\cos \pi nx$, where due to $q\in \cL_1$ a sequence
$(q_{c,n})_1^\iy\in \ell_1^2(\N)$. Then we obtain
$q(0)=q_{od}(0)+q_{ev}(0)$, since here all functions are continuous
on $[0,1]$. The first identity in \er{f3} is well known. Thus these
two facts imply \er{f3}. \BBox

We discuss trace formulas associated with the mappings $\t \star \m$
and $\gf$, defined by \er{dgf}. Here also we show \er{tm1},
\er{tm2}, since we can not find a reference for them.

 \begin{proposition}
\lb{Ttr1} Let $q\in \cL_1$. Then the trace formulas \er{tm1},
\er{tm2} and
\[
\lb{tr6}
\begin{aligned}
 \textstyle
   q(0)=\n_0+\sum_{n\ge1} \Big((\n_{n}-\m_n)+(\vr_{n}-\t_n)\Big),
   \\
 q(1)=\n_0+\sum_{n\ge1} \Big((\n_{n}-\m_n)-(\vr_{n}-\t_n)\Big),
 \end{aligned}
\]
and
\[
\lb{tr9}
\begin{aligned}
 \textstyle
   q(0)=2\sum_{n\ge1} \Big((\m_{n}^o-\m_n)+(\t_{n}^o-\t_n)\Big)
   =2\n_0+2\sum_{n\ge1} \Big((\n_{n}-\n_n^o)+(\vr_{n}-\vr_{n}^o)\Big),
   \\
 \textstyle   q(1)=2\sum_{n\ge1} \Big((\m_{n}^o-\m_n)-(\t_{n}^o-\t_n)\Big)
   =2\n_0+2\sum_{n\ge1} \Big((\n_{n}-\n_n^o)-(\vr_{n}-\vr_{n}^o)\Big).
 \end{aligned}
\]
 hold true, where all series converge absolutely.
 \end{proposition}

\no {\bf Proof.}  Firstly we discuss the trace  formulas for mixed
eigenvalues $\t_n$, since I can find their proofs. We show \er{tm1}
shortly repeating well-known arguments \cite{GL53}, \cite{Di63}. The
standard arguments from the papers of Gel'fand and Levitan
\cite{GL53} and Dikiy \cite{Di63} (see also a book \cite{L87}) imply
the identity
\[
\lb{tt1} \sum_{n\ge1} \big((\t_{n}^o-\t_n)+2\int_0^1q(x) \sin^2 k_n
xdx \big)=0,\qqq k_n=\sqrt{\t_n^o}>0,
\]
where $\sqrt 2\sin k_nx$ is the unperturbed eigenfunction
corresponding to the unperturbed eigenvalue $\t_n^o$. The identity
\er{f3} gives
$$-\sum_{n\ge1} \int_0^1q(x) 2\sin^2 k_nxdx=\sum_{n\ge1}
\int_0^1q\cos 2k_n x dx ={q(0)-q(1)\/4},
$$
which jointly with \er{tt1} yields \er{tm1}. The proof of \er{tm2}
for $\vr_n$ is similar.

Summing trace formulas in \er{tr1}, \er{tm1}, \er{tm2} we have the
first identity in \er{tr6}.

Summing trace formulas in \er{tr1}, summing trace formulas in
\er{tm1}, \er{tm2}
  and take their deferens   we obtain the second  identity in \er{tr6}.

Summing the first trace formula in \er{tr1} and \er{tm1}  and take
the deferens we have the first trace formula in \er{tr9}.

Summing the second trace formula in \er{tr1} and \er{tm2}  and take
the deferens we have the second trace formula in \er{tr9}.
 \BBox

\no $\bu $ Summing trace formulas in \er{tm1}, \er{tm2}  and take
the deferens we have
\[
\lb{tr7}
\begin{aligned}
 \textstyle
   \sum_{n\ge1} (\vr_{n}-\t_n)={q(0)-q(1)\/2},
   \qqq
   \sum_{n\ge1} (\vr_{n}+\t_n-2\t_n^o)=0.
 \end{aligned}
\]
Similar arguments and identities  \er{tr6}, \er{tr7} imply
\[
\lb{tr8}
\begin{aligned}
 \textstyle
\n_0+\sum_{n\ge1} (\n_{n}-\m_n)={q(0)+q(1)\/2},
   \qqq
\textstyle   \n_0+\sum_{n\ge1} (\n_{n}+\m_n-2\m_n^o)=0.
 \end{aligned}
\]

\setlength{\itemsep}{-\parskip} \footnotesize \no  {\bf
Acknowledgments.} I thank Segei Kuksin (Paris) for usefull discusion
about inverse problems.

\


\begin{thebibliography} {9999}\setlength{\itemsep}{-\parskip}
\footnotesize \footnotesize



\bibitem {CKK04} Chelkak, D.; Kargaev, P.; Korotyaev, E. Inverse problem
for harmonic oscillator perturbed by potential,
characterization. Comm. Math. Phys. 249 (2004), no. 1, 133--196.


\bibitem {CK09} Chelkak, D.; Korotyaev, E.  Weyl-Titchmarsh
functions of vector-valued Sturm-Liouville operators on the unit
interval. Journal of Functional Analysis, 257 (2009), 1546--1588.

\bibitem {CM17} Clay, M.; Margalit, D. Office Hours with a Geometric
Group Theorist. Princeton Univ. Press, 2017.



\bibitem{CM93} Coleman C.; McLaughlin, J. Solution of
the inverse problem for an impedance with integrable derivative, I
Commun. Pure Appl. Math. 46 (1993), 145--184.

\bibitem{CM93a} Coleman C.; McLaughlin, J. Solution of
the inverse problem for an impedance with integrable derivative II,
Commun. Pure Appl. Math. 46(1993), 185--212.

\bibitem {DT84}
 Dahlberg, B.; Trubowitz, E. The inverse Sturm-Liouville problem. III.
Comm. Pure Appl. Math. 37(1984), no. 2, 255--267.

\bibitem {Di63} Dikii, L. A.
On a formula of Gel'fand-Levitan. (Russian) Uspehi Matem. Nauk
(N.S.) 8, (1953). no. 2(54), 119--123.


\bibitem {GT87} Garnett, J., Trubowitz, E. Gaps and bands of one dimensional
periodic Schr\"odinger operators II. Comment. Math. Helv. 62(1987),
18--37.


\bibitem {GL51} Gel'fand, I. M.; Levitan, B. M. On the determination of a differential equation
from its spectral function. (Russian) Izvestiya Akad. Nauk SSSR.
Ser. Mat. 15, (1951). 309--360. English Translation: Amer. Math.
Soc. Transl. (2) 1 (1955), 253--304.


\bibitem {GL53}
 Gel'fand, I. M.; Levitan, B. M. On a simple identity for the characteristic values of a differential
 operator of the second order. (Russian) Doklady Akad. Nauk SSSR (N.S.) 88, (1953). 593--596.

\bibitem {GR88} Guillot J.G.; Ralston J.V. Inverse spectral theory for a
singular Sturm-Liouville operator on $[0,1]$, J. Diff. Eq., 76
(1988), 353--373.

\bibitem {IT83}
Isaacson, E. L.; Trubowitz, E. The inverse Sturm-Liouville problem.
 I. Comm. Pure Appl. Math. 36 (1983), no. 6,   767--783.

\bibitem {IMT84} Isaacson, E. L.; McKean, H. P.; Trubowitz, E.
The inverse Sturm-Liouville problem. II. Comm. Pure Appl. Math. 37
(1984), no. 1, 1--11.


\bibitem {IM75} Its, A. R.; Matveev, V. B. Schr\"odinger operators with the finite-band
spectrum and the N-soliton solutions of the Korteweg-de Vries
equation. (Russian) Teoret. Mat. Fiz. 23 (1975), no. 1, 51--68.

\bibitem {KP03} Kappeler, T.; P\"oschel, J. KdV $\&$ KAM. Springer, 2003.


\bibitem {KK97} Kargaev, P. ;  Korotyaev E.
The inverse problem for the Hill operator, direct approach. Invent.
Math.  129(1997), 567--593.

\bibitem {KK00} Klein, M.; Korotyaev, E. Parametrization of periodic
weighted operators in terms of gap lengths. Inverse Problems 16
(2000), no. 6, 1839--1860.

\bibitem {K19} Korotyaev, E. Inverse Sturm-Liouville problems for
non-Borg conditions,  Journal of Inverse and Ill-Posed Problems,
27(2019), no 3. 445--452.

\bibitem {K06}  Korotyaev, E. Estimates for the Hill operator. II. J.
Differential Equations 223 (2006), no. 2, 229--260.

\bibitem {K00jde}   Korotyaev, E. Estimates for the Hill operator. I. J.
Differential Equations 162 (2000), no. 1, 1--26.

\bibitem {K00}  Korotyaev, E Inverse problem for periodic "weighted"
operators. J. Funct. Anal. 170 (2000), no. 1, 188--218.

 \bibitem {K99}  Korotyaev, E. The inverse problem and trace
formula for the Hill operator, II.  Math. Z. 231(1999), 345--368.

\bibitem {K98} Korotyaev, E. Estimates of periodic potentials in terms
of gap lengths. Comm. Math. Phys. 197 (1998), no. 3, 521--526.


\bibitem {K97} Korotyaev, E. The inverse problem for the Hill operator.
I. Int. Math. Res. Notices 1997, no. 3, 113--125.


\bibitem {K97m}  Korotyaev, E. The estimates of periodic potentials in
terms of effective masses. Comm. Math. Phys. 183 (1997), no. 2,
383--400.

\bibitem {KC09}  Korotyaev, E.; Chelkak,  D.
 The inverse Sturm-Liouville problem with mixed boundary conditions,
St. Petersburg Math.  Journal. 21(2009), no 5, 114--137.

\bibitem {Kr51}
Krein, M. G. Solution of the inverse Sturm-Liouville problem.
(Russian) Doklady Akad. Nauk SSSR (N.S.) 76, (1951). 21--24.


\bibitem {Kr54}
Krein, M. G. On a method of effective solution of an inverse
boundary problem. (Russian) Doklady Akad. Nauk SSSR (N.S.) 94,
(1954). 987--990.

\bibitem{L87} Levitan, B. Inverse Sturm-Liouville problems.
Utrecht: VNU Science Press, 1987.

\bibitem {MW66} W. Magnus, S. Winkler. Hill's equation,
Interscience Tracts in Pure and Applied Mathematics,
No. 20 Interscience Publishers John Wiley $\&$ Sons, New
York-London-Sydney, 1966.

\bibitem {M50}  Marchenko, V. A. Concerning the theory of a differential operator
of the second order. (Russian) Doklady Akad. Nauk SSSR. (N.S.) 72,
(1950). 457--460.


\bibitem {MO75} Marchenko, V.; Ostrovski, I. A
characterization of the spectrum
 of the Hill operator. Mat. Sb. 97(139), (1975), 540--606.

 \bibitem{M86}
 Marchenko, V. Sturm-Liouville operator and applications.
Basel, Birkh\"auser 1986.


\bibitem{PT87} P\"oschel, P., Trubowitz E. Inverse Spectral Theory.
Boston, Academic Press, 1987.


\bibitem{SS08}  Savchuk, A. M.; Shkalikov, A. A.
On the properties of mappings associated with inverse
  Sturm-Liouville problems.  Proc. Steklov Inst. Math. 260 (2008), no. 1,
  218--237.

  \bibitem{SS10} Savchuk, A. M.; Shkalikov, A. A.
Inverse Problems for Sturm-Liouville Operators with Potentials in
Sobolev Spaces:
  Uniform Stability, Funct. Anal. Appl., 44:4 (2010), 270--285.

\bibitem{T77} Trubowitz, E. The inverse problem for periodic potentials.
 Comm. Pure Appl. Math. 30 (1977), no. 3, 321--337.

\end{thebibliography}
\end{document}